# A Canonical Quantization formalism of curvature squared action

Mahgoub salih
Department of physics
King Saud university
Teachers college
msalih@ksu.du.sa

Abstract

the generalized Einstein action is treated quantum mechanically by using a quadratic lagrangian form. The canonical quantization of this action is obtained by using the auxiliary variable to define the generalized momentum. Physical constraints are imposed on the surface term, which is defined to be the cosmological constant. One obtains the familiar Wheeler-de Witt equation. The solution of this equation is in conformity with General Relativity (GR) .In addition to the fact that it is free from GR setbacks at the early universe, since it gives time decaying cosmological constant.

The wave function of the universe and the cosmic scale factor are complex quantities, which indicates the existance of quantum effects.2.7k ,CBR temperature is calculated .

## *1.Hamiltonian of curvature squared action:*

Palatini formalism which is based on independent variations of the metric and the connection, has been known to be equivalent to the corresponding metric formulation for the lagrangian which depends linearly on the scalar curvature constructed out of the metric and the Ricci tensor of the connection .In this case, in fact, field equations imply that the connection has to be the Levi-Civita connection of the metric, which in turn satisfies Einstein's equations[1].This means that there is a full dynamical equivalence with the corresponding metric lagrangian , which a posteriori coincides with the Hilbert lagrangian of general relativity .

A generic fourth –order theory in four dimensions can be described by the action [2]

$$A = \int_M f(R)\sqrt{-g}d^n x \qquad 1-1$$

Where $f(R)$is a function of Ricci scalar R.

The field equations are:

$$\acute{f}(R)R_{\alpha\beta} - \frac{1}{2}f(R)g_{\alpha\beta} = \acute{f}(R)^{;\alpha\beta}\left(g_{\alpha\mu}g_{\beta\nu} - g_{\alpha\beta}g_{\mu\nu}\right) \qquad 1-2$$

This can be expressed in the more expressive form

$$G_{\alpha\beta} = R_{\alpha\beta} - \frac{1}{2}g_{\alpha\beta}R = T_{\alpha\beta} \qquad 1-3$$

$$T_{\alpha\beta} = \frac{1}{\acute{f}(R)}\left[\frac{1}{2}g_{\alpha\beta}\left(f(R) - R\acute{f}(R) + \acute{f}(R)^{;\alpha\beta}\left(g_{\alpha\mu}g_{\beta\nu} - g_{\alpha\beta}g_{\mu\nu}\right)\right]\right. \qquad 1-4$$

Where $T_{\alpha\beta}$ is the curvature-energy tensor. The prime indicates the derivative with respect to R.

It is possible to reduce the action to a point-like, Friedmann-Robertson-walker(F R W) one. We have to write

$$A = \int dt L\left(a, \dot{a}, R, \dot{R}\right) \qquad 1-5$$

Where dot indicates derivative with respect to the cosmic time. The scale factor (a) and the Ricci scalar R are the canonical variables. This position could seem arbitrary since R depends on $a, \dot{a}, \ddot{a}$.The definition of R in terms of $a, \dot{a}, \ddot{a}$ introduces a constraint which eliminates second and higher order derivatives in action(1-5),and gives a system of second order differential equations in $\{R, a\}$.

Action (1-5) can be written as:

$$A = \int dt \left\{a^3 f(R) - \lambda \left[R + 6\left(\frac{\ddot{a}}{a} + \frac{\dot{a}^2}{a^2} + \frac{k}{a^2}\right)\right]\right\} \qquad 1-6$$

Where the lagrangian multiplier $\lambda$ is derived by varying action (1-6) with respect to R .this yield

$$\lambda = a^3 \grave{f}(R)$$

If we consider homogeneous and isotropic Robertson-Walker metric in form of

$$dS^2 = g_{\mu\upsilon} dX^\mu dX^\upsilon = dt^2 - a^2(t)\left[\frac{dr^2}{1-kr^2} + r^2(d\theta^2 + sin^2\theta d\phi^2)\right] \qquad 1-7$$

For which

$$R_{ij} = \left(\frac{\ddot{a}}{a} + \frac{\dot{a}^2}{a^2} + \frac{k}{a^2}\right) g_{ij} \qquad 1-8$$

$$R = g^{ij} R_{ij}$$

Then scalar curvature R will be written as

$$R = 6\left(\frac{\ddot{a}}{a} + \frac{\dot{a}^2}{a^2} + \frac{k}{a^2}\right) \qquad 1-9$$

If we assume that $f(R) = R^{\frac{l}{2}}$, action (1-6) will be written as:

$$A = 2(1-n)\pi^2\beta \int a^3 R^{\frac{l}{2}} dt \qquad 1-10$$

If we assume $l = 4$,

$$A = -216\pi^2\beta \int a^3 \left(\frac{\ddot{a}}{a} + \frac{\dot{a}^2}{a^2} + \frac{k}{a^2}\right)^2 dt \qquad 1-11$$

If we put $m = -216\pi^2\beta$

$$A = m \int \left(a\ddot{a}^2 + 2\ddot{a}(\dot{a}^2 + k) + \frac{k^2}{a} + \frac{\dot{a}^4}{a} + 2k\frac{\dot{a}^2}{a}\right) dt \qquad 1-12$$

The corresponding Lagrangian in a minisuperspace model involves field variables along with their first derivatives and second derivatives. Therefore, in order to cast the action in canonical form that demands the corresponding Lagrangian be expressed in terms of the field variables along with their first derivatives only, one requires to introduce auxiliary variables[3,4,5,6,7,8,9]. It has been observed that any form of auxiliary variable produces correct classical field equations in homogeneous and isotropic model. Further, it is not enough to get the correct classical field equations only, rather one should also be able to produce correct and well behaved, quantum description of the system under consideration. By the above statement we mean to find a variable, if it at all exists, with respect to which the Hamiltonian would turn out to be hermitian.

Starobinsky [10],was The first one who established the relevance of the fourth order gravity theory in cosmology though it appeared earlier in the context of quantum field theory in curved space time. Starobinsky [10] presented a solution of the inflationary scenario without invoking phase transition in the very early Universe, from a field equation containing only the geometric terms. However, the field equations could not be obtained from the action principle, as the terms in the field equations are generated from the perturbative quantum field theory. Later, Starobinsky and Schmidt [11] have shown that the inflationary phase is an attractor in the neighborhood of the fourth order gravity theory.

Quantum cosmology is most elusive as the role of time is not unique and one is being unable to define the Hilbert space[12]. This is due to the fact that 'time' is not an external parameter in general theory of relativity, rather it is intrinsically contained in the theory, unlike its role in quantum mechanics or quantum field theory in flat space time. In curved space time, different slices correspond to different choices of time leading to inequivalent quantum theories. Likewise, the canonical quantization of gravity is devoid of an unique time variable and hence the definition of probability of emergence of a particular Universe out of an ensemble is ambiguous. The canonical quantization of Einstein-Hilbert action together with some matter fields yields the Wheeler-deWitt equation which does not contain time a priori, although it emerges intrinsically through the scale factor of the Universe. However, if the canonical variable is so chosen that one of the true degrees of freedom is disentangled from the kinetic part of the canonical variables, then this kinetic part in the corresponding quantum theory yields a quantum mechanical flavor of time. This is possible only if the Einstein-Hilbert action is replaced by curvature squared action or modified by the introduction of curvature squared term, in the Robertson-Walker minisuperspace model.

Such wonderful relevance of higher order gravity theory in the context of cosmology inspired some authors to give interpretation of the quantum cosmological wavefunction with $R^2$ [13] and even $R^3$ [14] terms in the Einstein-Hilbert action.

Further the functional integral for the wavefunction of the Universe proposed by Hartle-Hawking [15] runs into serious problem, since the wavefunction diverges badly. There are some prescriptions [16] to avoid such divergences, though a completely satisfactory result has not yet been obtained. However, to get a convergent functional integral,Horowitz [9] proposed an action in the form

$$S = \int d^4X\sqrt{g}\left[AC_{ijkl}^2 + B(R - 4\lambda)^2\right] \qquad 1-13$$

Where $C_{ijkl}^2$ is the Weyl tensor, R is the Ricci scalar, $\lambda$ is the cosmological constant and A,B are the coupling constants.

The action (1-13 ) reduces to the Einstein-Hilbert action at the weak energy limit. To obtain a workable and simplified form of the field equations, one may consider a spatially homogeneous and isotropic minisuperspace background, for which the Weyl tensor trivially vanishes. A.K.Sanyal and B.Modak [3], considered the action like (1-11) retaining only the curvature squared term. The field equations for such an action can be obtained by the standard variation principle.

In the variational principle, the total derivative terms in the action are extracted and one gets a surface integral which is assumed to vanish at the boundary or the action is chosen in such a way that those surface integral terms have no contribution. However, for canonical quantization this principle is not of much help and one has to express the action in the canonical form, which is achieved only through the introduction of the auxiliary variable. Auxiliary variable can be chosen in an adhoc manner, and different choice of such variable would lead to different description of quantum dynamics, keeping the classical field equations unchanged. In view of this Ostrogradski [7] in one hand and Boulware etal [6] on the other, have made definite prescriptions to choose such variables. Ostrogradski's prescription [7] was followed by Schmidt [17].

in Boulware etal's [6] prescription the auxiliary variable should be chosen by taking the first derivative of the action with respect to the highest derivative of the field variable present in the action.

Hawking and Luttrell [18] utilized Boulware etal's [6] technique to identify the new variable and showed that the Einstein-Hilbert action along with a curvature squared term reduces to the Einstein-Hilbert action coupled to a massive scalar field, assuming the conformal factor. Horowitz on the other hand, showed that the canonical quantization of the curvature squared action yields an equation which is similar to the Schrodinger equation[9]. Pollock [19] also used the same technique to the induced theory of gravity and obtained the same type of result as that obtained by Horowitz [9], in the sense that the corresponding Wheeler-de Witt equation looks similar to the Schrodinger equation.

The striking feature of Boulware etal's [6] prescription is that it can even be applied in situations where introduction of the auxiliary variable is not at all required, e.g. in the induced theory of gravity, vacuum Einstein-Hilbert action etc.

The classical field equations remain unchanged with or without the introduction of the auxiliary variables.

***2 Hamiltonian with respect to ( ä):***

Define the auxiliary variable according to Horowitz [9]

$$Q^{ab} = \pi^{ab} = \frac{\partial A}{\partial\left(\frac{\partial h_{ab}}{\partial t}\right)} \qquad 2-1$$

Where $h_{ab}$ is the three metric.

Action(1-5) can be written in the following form:

$$A = m\int\left(a\ddot{a}^2 + 2\ddot{a}(\dot{a}^2 + k) + \frac{k^2}{a} + \frac{\dot{a}^4}{a} + 2k\frac{\dot{a}^2}{a}\right)dt \qquad 2-2$$

Where the Lagrange is

$$L = m\left(a\ddot{a}^2 + 2\ddot{a}(\dot{a}^2 + k) + \frac{k^2}{a} + \frac{\dot{a}^4}{a} + 2k\frac{\dot{a}^2}{a}\right) \qquad 2-3$$

The auxiliary variable Q is defined as the variation of the action with respect to the higher derivative term $\ddot{a}$ ;

$$Q_1 = \frac{\partial A}{\partial \ddot{a}} = 2m\left(\frac{\ddot{a}}{a} + \frac{\dot{a}^2}{a^2} + \frac{k}{a^2}\right)a^2 \qquad 2-4$$

According to[3], every action should be supplemented by an appropriate boundary term. Here we rid of all the zero-order derivative terms to the introduction of auxiliary variable.

The zero-order derivative term in action (1-16 ) is $\frac{k^2}{a}$.

Then action (1-16 ) is written in the following form:

$$A = m\int\left(a\ddot{a}^2 + 2\ddot{a}(\dot{a}^2 + k) + \frac{\dot{a}^4}{a} + 2k\frac{\dot{a}^2}{a}\right)dt + m\int\frac{k^2}{a}dt \qquad 2-5$$

Now we can write action ( 2-2) in canonical form as:

$$A=\int\left(\frac{Q_1^2}{4ma}-m\frac{k^2}{a}\right)dt+m\int\frac{k^2}{a}dt \qquad 2-6$$

$$A=\int\left(-\dot{Q}_1\dot{\gamma}+\frac{Q_1^2}{4ma}-m\frac{k^2}{a}\right)dt+\int\dot{Q}_1\dot{\gamma}dt+\sigma \qquad 2-7$$

$$\dot{\gamma}=-\dot{a}-\int\frac{\dot{a}^2+k}{a}dt \qquad 2-8$$

$$\sigma=m\int\frac{k^2}{a}dt \qquad 2-9$$

$$\Sigma_1=\int\dot{Q}_1\dot{\gamma}dt$$

Finally, the action became in canonical form with the suitable boundary term like

$$A=\int\left(-\dot{Q}_1\dot{\gamma}+\frac{Q_1^2}{4ma}-m\frac{k^2}{a}\right)dt+\Sigma_1+\sigma \qquad 2-10$$

Assuming all boundary terms vanish at boundary $\Sigma_1+\sigma=0$. This leads to:
$k=0$,and the scale factor (a)is a constant.
Now we apply Eular-lagrange equation ($\frac{d}{dt}\frac{\partial L}{\partial \dot{Q}}-\frac{\partial L}{\partial Q}=0$ ) to the action to recover definition of the auxiliary variable

$$Q_1=2ma\ddot{a}+2m(\dot{a}^2+k)=2m\left(\frac{\ddot{a}}{a}+\frac{\dot{a}^2}{a^2}+\frac{k}{a^2}\right)a^2 \qquad 2-11$$

This is the same definition of the auxiliary as it was defined before.
Now it will be easy to obtain the Hamiltonian according to:

$$H(q,p,t)=\dot{q}_ip_i-L(q,\dot{q},t)$$

$$p_{Q_1}=\frac{\partial L}{\partial\dot{Q}_1}=-\dot{\gamma}$$

$$p_\gamma=\frac{\partial L}{\partial\dot{\gamma}}=-\dot{Q}_1$$

$$H=\dot{\gamma}p_\gamma-\frac{{Q_1}^2}{4ma}+\frac{mk^2}{a} \qquad 2-12$$

For canonical quantization we express the Hamiltonian in terms of the basic variables, for which we choose

$$p_x=Q_1=\frac{\partial L}{\dot{x}}$$

$$x=\dot{a}$$

$$p_{Q_1}=-\dot{\gamma}$$

then we obtain the Hamiltonian in terms of the basic variables as:

$$H=\dot{\gamma}p_\gamma-\frac{{p_x}^2}{4ma}+\frac{mk^2}{a} \qquad 2-13$$

As we know $H=0$ then;

$$-\dot{\gamma}p_\gamma=-\frac{{p_x}^2}{4ma}+\frac{mk^2}{a}$$

To get the dynamical equation( $p=-i\hbar\nabla$ )

$$i\hbar\frac{\partial}{\partial\gamma}=\frac{\hbar^2}{4ma\dot{\gamma}}\frac{\partial^2}{\partial x^2}+\frac{mk^2}{a\dot{\gamma}} \qquad 2-14$$

This can be expressed as: $i\hbar\frac{\partial\psi}{\partial a}=\widehat{H_1}\psi$
Where: $\widehat{H}$ is the effective Hamiltonian.

$$\widehat{H_1} = \frac{\hbar^2}{4ma\dot{\gamma}}\frac{\partial^2}{\partial x^2} + V_{e1} \qquad 2-15$$

$$V_{e1} = \frac{mk^2}{a\dot{\gamma}} \qquad 2-16$$

Represents the effective potential.

Assuming perfect fluid, which obey the equation of state $p = \omega\rho$ now we can solve for $\rho$ *and* $\dot{\rho}$ in equation (3-9) and with aid of equation $\frac{\ddot{a}}{a} + 2\left(\frac{\dot{a}}{a}\right)^2 + 2\frac{k}{a^2} = 4\pi G(\rho - p)$.

If the potential (2-16) remains constant with respect to time (t).we found the relation of energy density and the scale factor ,

$$\frac{\partial}{\partial t}\left(\frac{mk^2}{a\dot{\gamma}}\right) = 0$$

$$\dot{a}\dot{\gamma} + a\ddot{\gamma} = 0$$

$$\dot{a}^2 + \dot{a}\int\frac{\dot{a}^2 + k}{a}dt + a\ddot{a} + \dot{a}^2 + k = 0$$

Add k and divide by $a^2$, after all we find:

$$\frac{2\dot{a}^2}{a^2} + \frac{2k}{a^2} + \frac{\ddot{a}}{a} = -\frac{\dot{a}}{a^2}\int\frac{\dot{a}^2 + k}{a}dt + \frac{k}{a^2} \qquad 2-17$$

Now we can substitute equation $(2-17)$ into $\left(\frac{\ddot{a}}{a} + 2\left(\frac{\dot{a}}{a}\right)^2 + 2\frac{k}{a^2} = 4\pi G(\rho - p)\right)$

$$4\pi G\rho(1-\omega) = -\frac{\dot{a}}{a^2}\int\frac{\dot{a}^2 + k}{a}dt + \frac{k}{a^2} \qquad 2-18$$

$$\dot{\rho} = \frac{\partial\rho}{\partial t}$$

$$4\pi G\dot{\rho}(1-\omega) = -\frac{\dot{a}^2}{a^3}\left[\dot{a} + \left(\frac{a\ddot{a}}{\dot{a}^2} - 2\right)\int\frac{\dot{a}^2 + k}{a}dt + \frac{3k}{\dot{a}}\right] \qquad 2-19$$

$$\frac{\dot{\rho}}{\rho} = -\frac{\dot{a}}{a}\left[\frac{\dot{a} + \left(\frac{a\ddot{a}}{\dot{a}^2} - 2\right)\int\frac{\dot{a}^2+k}{a}dt + \frac{3k}{\dot{a}}}{-\int\frac{\dot{a}^2+k}{a}dt + \frac{k}{\dot{a}}}\right] \qquad 2-20$$

$$\omega_1 = \frac{1}{3}\left[\frac{\dot{a} + \left(\frac{a\ddot{a}}{\dot{a}^2} - 2\right)\int\frac{\dot{a}^2+k}{a}dt + \frac{3k}{\dot{a}}}{-\int\frac{\dot{a}^2+k}{a}dt + \frac{k}{\dot{a}}}\right] - 1 \qquad 2-21$$

we consider two cases as:

1. $a \to \infty; \dot{a} \ll 1; k \neq 0$ ,

This implies that

$$\frac{\dot{\rho}}{\rho} \sim -3\frac{\dot{a}}{a} \qquad 2-22$$

or

$$\rho \propto a^{-3} \qquad 2-23$$

$$p \sim 0 \qquad 2-24$$

with

$$\omega_1 \sim 0 \qquad 2-25$$

This represent matter dominated universe, this means that our model described matter era at large radius and slow expansion rate, also we found that the density is decreasing with increasing of the radius in conformity with GR and experiments.

2. $a \to 0; \dot{a} \to \infty; k \neq 0$

$$\frac{\dot{\rho}}{\rho} \sim -2\frac{\dot{a}}{a} \qquad 2-26$$

where

$$\omega_1 \sim -\frac{1}{3} \qquad 2-27$$

which is requires negative pressure

$$p \sim -\frac{1}{3}\rho \qquad 2-28$$

This result represents early universe ,which predict inflation.

### *3. Hamiltonian with respect to ( $\dot{a}^2$):*

Now according to Boulware's prescription[6], Hamiltonian formulation of an action containing higher order curvature invariant term requires introduction of auxiliary variable,

$$Q^{ab} = \pi^{ab} = \frac{\partial A}{\partial\left(\frac{\partial k_{ab}}{\partial t}\right)}$$

And the extrinsic curvature $k_{ab}$ is:

$$k_{11} = k_{22} = k_{33} = -a\dot{a} \qquad 3-1$$

The action of $R^2$ can be written as:

$$A = m\int\left(a\ddot{a}^2 + 2\ddot{a}(\dot{a}^2 + k) + \frac{k^2}{a} + \frac{\dot{a}^4}{a} + 2k\frac{\dot{a}^2}{a}\right)dt \qquad 3-2$$

If we get rid the zero-order derivative $\frac{k^2}{a}$ to auxiliary variable [4],

$$A = m\int\left(a\ddot{a}^2 + 2\ddot{a}(\dot{a}^2 + k) + \frac{\dot{a}^4}{a} + 2k\frac{\dot{a}^2}{a}\right)dt + m\int\frac{k^2}{a}dt \qquad 3-3$$

$$A = m\int\left(a\ddot{a}^2 + 2\ddot{a}(\dot{a}^2 + k) + \frac{\dot{a}^4}{a} + 2k\frac{\dot{a}^2}{a}\right)dt + \sigma \qquad 3-4$$

Now we supplement the action with boundary term

$$\sigma = m\int\frac{k^2}{a}dt$$

This action can be written in the following canonical form, with

$$Q_2 = \frac{\partial A}{\dot{a}^2} = 2m\left(\frac{\ddot{a}}{a} + \frac{\dot{a}^2}{a^2} + \frac{k}{a^2}\right)a \qquad 3-5$$

$$A = \int\left(-\dot{Q}_2\dot{\alpha} + \frac{a{Q_2}^2}{4m} - m\frac{k^2}{a}\right)dt + \Sigma_2 + \sigma \qquad 3-6$$

$\Sigma_2 = \int \dot{Q}_2\dot{\alpha}$

$$\dot{\alpha} = -a\dot{a} - \int(k)dt \qquad 3-7$$

Assuming all boundary terms vanishing at boundary $\Sigma_2 + \sigma = 0$ . This leads to the same result as before:

$k = 0$,and the scale factor (a)is a constant.

As before we recovered the definition of the auxiliary variable by applying Eular-lagrange.

For canonical quantization, we express the Hamiltonian in terms of the basic variables, for which we choose:

$$\acute{x} = \int \dot{a}^2 dt$$
$$\dot{\acute{x}} = \dot{a}^2$$

and

$$Q_2 = p_{\acute{x}}$$

Now we obtain the Hamiltonian($H = 0$) as:

$$-\dot{\alpha} p_\alpha = -\frac{a {Q_2}^2}{4m} + \frac{mk^2}{a} \qquad 3-8$$

We express the dynamical equation as:

$$i\hbar \frac{\partial}{\partial \alpha} = \frac{a\hbar^2}{4m\dot{\alpha}} \frac{\partial^2}{\partial \acute{x}^2} + \frac{mk^2}{a\dot{\alpha}} \qquad 3-9$$

$$\widehat{H_2} = \frac{a\hbar^2}{4m\dot{\alpha}} \frac{\partial^2}{\partial \acute{x}^2} + V_{e2} \qquad 3-10$$

where:

$$V_{e2} = \frac{mk^2}{a\dot{\alpha}} \qquad 3-11$$

$$4\pi G \dot{\rho}(1-\omega) = -\left\{\frac{\dot{a}^2}{a^4}\right\}\left[\left(\frac{\ddot{a}a}{\dot{a}^2} - 3\right)\int kdt + 3k\frac{a}{\dot{a}}\right] \qquad 3-12$$

$$4\pi G \rho(1-\omega) = \left\{\frac{\dot{a}}{a^3}\right\}\left[-\int kdt + k\frac{a}{\dot{a}}\right] \qquad 3-13$$

$$\frac{\dot{\rho}}{\rho} = -\frac{\dot{a}}{a}\left\{\frac{\left(\frac{\ddot{a}a}{\dot{a}^2} - 3\right)\int kdt + 3k\frac{a}{\dot{a}}}{-\int kdt + k\frac{a}{\dot{a}}}\right\} \qquad 3-14$$

$$\omega_2 = \frac{1}{3}\left[\frac{\left(\frac{\ddot{a}a}{\dot{a}^2} - 3\right)\int kdt + 3k\frac{a}{\dot{a}}}{-\int kdt + k\frac{a}{\dot{a}}}\right] - 1 \qquad 3-15$$

According to relevant treatment of the potential, also, there are tow cases, but they are equivalent :

1. $a \to \infty; \dot{a} \ll 1; k \neq 0$

$$\frac{\dot{\rho}}{\rho} \sim -3\frac{\dot{a}}{a}$$

this implies that

$$\omega_2 \sim 0 \qquad 3-16$$

and

$$p \sim 0$$
$$\rho \propto a^{-3}$$

2. $a \to 0; \dot{a} \to \infty; k \neq 0$

$$\frac{\dot{\rho}}{\rho} \sim -3\frac{\dot{a}}{a}$$

this implies that

$$\omega_2 \sim 0 \qquad 3-17$$

and

$$p \sim 0$$
$$\rho \propto a^{-3}$$

***4. Comparison with A.K Sanyal work:***

A.K Sanyal (according to Boulware`s prescription) choose a variable $z = \frac{a^2}{2}$ then he found the definition of the auxiliary variable Q as:

$$Q = \frac{\partial A}{\partial \ddot{z}} = 144\pi^2\beta\left(\frac{\ddot{a}}{a} + \frac{\dot{a}^2}{a^2} + \frac{k}{a^2}\right)a$$

Then Sanyal suggestion is to get rid of all the total derivative terms prior to the introduction of auxiliary variable. After doing this, he found the boundary term that supplements the action is different from that one which supplements the action in references [3,4,5] when he followed Horowitz[9].A particular action must be supplemented by the same boundary term, independent of the choice of variable.

we found the same boundary terms in this work according to both Horowitz and Boulware`s prescription, $\sigma = m\int \frac{k^2}{a} dt$

At boundaries we found $k = 0$,and the scale factor (a)is a constant.

***5. Acceleration and deceleration scenario:***

In this model we did not include matter term in action (1-12),so, this model represents curvature effects in all universal parameters, which will be added to that parameters which was obtained from matter lagrangian.

The total effective Hamiltonian of $R^2$ in our minisuperspace model is :

$$\hat{H} = \hat{H}_1 + \hat{H}_2 = \frac{\hbar^2}{4ma\dot{\gamma}}\frac{\partial^2}{\partial x^2} + \frac{a\hbar^2}{4m\dot{\alpha}}\frac{\partial^2}{\partial \acute{x}^2} + V \qquad 5-1$$

$$V = V_{e1} + V_{e2} = \frac{mk^2}{a}\left[\frac{1}{\dot{\gamma}} + \frac{1}{\dot{\alpha}}\right] \qquad 5-2$$

The wave function is:

$$\psi = \psi_1 + \psi_2 = \mathrm{C}_1\mathrm{e}^{\frac{2\mathrm{imkx}}{\hbar}} + \mathrm{C}_2\mathrm{e}^{\frac{-2\mathrm{imkx}}{\hbar}} + D_1 e^{\frac{2ikm\,\acute{x}}{a\hbar}} + D_2 e^{\frac{-2ikm\,\acute{x}}{a\hbar}} \qquad 5-3$$

$$\omega = \omega_1 + \omega_2$$

$$\omega = \frac{\dot{a} + \left(\frac{a\ddot{a}}{\dot{a}^2} - 2\right)\int\frac{\dot{a}^2+k}{a}dt + \frac{3k}{\dot{a}}}{-3\int\frac{\dot{a}^2+k}{a}dt + 3\frac{k}{\dot{a}}} + \frac{\left(\frac{\ddot{a}a}{3\dot{a}^2} - 1\right)\int k dt + k\frac{a}{\dot{a}}}{-\int k\,dt + k\frac{a}{\dot{a}}} - 2 \qquad 5-4a$$

$$\omega = \frac{\dot{a} - (q+2)B + 3\frac{k}{\dot{a}}}{3\frac{k}{\dot{a}} - 3B} + \frac{\left(\frac{q}{3} + 1\right)t - H^{-1}}{t - H^{-1}} - 2 \qquad 5-4b$$

Where $q = -\frac{a\ddot{a}}{\dot{a}^2}$ ,the deceleration parameter, H is Hubble parameter $H = \frac{\dot{a}}{a}$ ,$B = \int\frac{\dot{a}^2+k}{a}dt$

There are two approaches, first of them represents the early universe:

1. $a \to 0$;

   at this early time we found that, $\omega$, is dominated by the term: $\dot{a} - 2\int\frac{\dot{a}^2}{a}dt$ ,to study how this term leads to acceleration universe, write friedmann equation:

$$\frac{\ddot{a}}{a} = -\frac{4\pi G}{3}\rho(1 + 3\omega)$$

- If $\dot{a} \ll \int\frac{\dot{a}^2}{a}dt$ ,this case happen when we assume that the universe has phase transition, which change $\dot{a}$ from zero to infinity. we can rewrite the integral;

$$\int \frac{\dot{a}^2}{a} dt = \int \frac{\dot{a}}{a} da$$

now we find the upper critical value of $\omega$:

$\omega \to -\frac{1}{3}$ , and, $\ddot{a} \to 0$ .this means that, curvature did not involve in universe revolution.

- If $\dot{a}$ is less than but comparable with $\int \frac{\dot{a}^2}{a} dt$ , we find:

  $\omega < -\frac{1}{3}$ ,and, $\ddot{a} > 0$. this implies negative pressure, $p < 0$ ,and $\rho > a^{-2}$

The negative pressure leads to inflation [20].

Now if we let $a \propto t^n$,we find the following relation:

$$\omega = \frac{3 - 4n}{3n} \qquad n \neq 0 \qquad 5-5$$

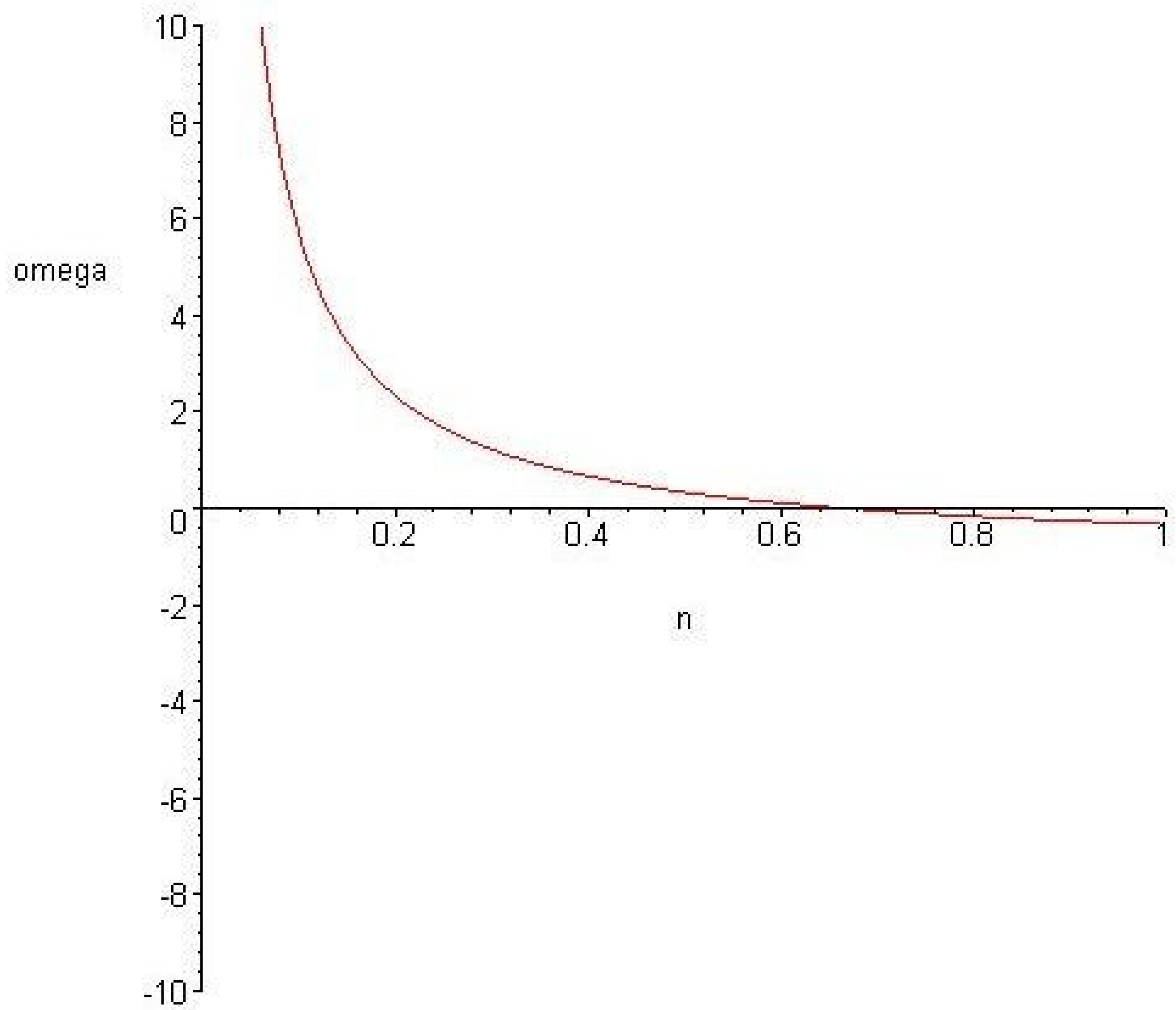

fig$(5-1)$variation of ω with respect to n, $a \to 0$

From above equation when

- $|n| \gg 1$ , $\omega \to -\frac{4}{3}$ .
- $|n| \ll 1$, $\omega \to \frac{1}{n}$ .

then, $$-\frac{4}{3} \leq \omega \leq \frac{1}{|n|}$$ .

2. Second approach :$a \to \infty; \dot{a} \ll 1$;

- $\frac{\ddot{a}a}{\dot{a}^2} \approx 0$ ,this yields one value of $\omega$, where $\omega = 0$ , and, $\ddot{a} < 0$ ,$\rho \propto a^{-3}$ ,$p = 0$,this represents matter(relativistic particles) ,

If we turn back to action (1-6), adding matter lagrangian, means, adding a pressure term [ 21].So, according to second approach the universe will contract very fast. May be back to the beginning and, pre-big bang (oscillating).

***6. Cosmological constant:***

The boundary terms in equation (2-10) and (3-6) represent the cosmological constant.

$$\Lambda_1 = \dot{Q}_1\dot{\gamma} + m\frac{k^2}{a} \qquad 6-1a$$

$$\Lambda_1 = m\frac{k^2}{a} - 2m\left(\dot{a} + \int \frac{\dot{a}^2 + k}{a} dt\right)(3\dot{a}\ddot{a} + a\dddot{a}) \qquad 6-1b$$

$$\dot{Q}_1 = 2m(3\dot{a}\ddot{a} + a\dddot{a})$$

If the cosmological constant equal zero:

$$\Lambda_1 = 0$$

$$\dot{Q}_1\dot{\gamma} + m\frac{k^2}{a} = 0$$

For simplicity we let

$$k = 0$$

In this case

$$\left(-\dot{a} - \int \frac{\dot{a}^2}{a} dt\right)(3\dot{a}\ddot{a} + a\dddot{a}) = 0 \qquad 6-2$$

Solving equation (6-2)

$$3\dot{a}\ddot{a} + a\dddot{a} = 0 \qquad 6-3$$

$$\dot{a} + \int \frac{\dot{a}^2}{a} dt = 0 \qquad 6-4$$

integrate equation (6-3) :

$$a\ddot{a} + \dot{a}^2 = C$$

$$a\frac{d^2a}{dt^2} + \left(\frac{da}{dt}\right)^2 + C = 0 \qquad 6-5$$

Differentiate equation (6-4):

$$\frac{d}{dt}\left(\dot{a} + \int \frac{\dot{a}^2}{a} dt\right) = 0$$

$$a\ddot{a} + \dot{a}^2 = 0$$

$$a\frac{d^2a}{dt^2} + \left(\frac{da}{dt}\right)^2 = 0 \qquad 6-6$$

If the integration constant $C = 0$,equation (6-5) construe to equation (6-6).

Solving equation (6-6),if $a \propto t^n$or $a = At^n$,A is a constant.

We find that:

$$t^{2n-2}[(n-1) + n] = 0$$

$$n = \frac{1}{2}$$

By substituting in equation( 4.3-1b) that $a = At^n$ we find:

$$\int \frac{\dot{a}^2+k}{a} dt = \frac{An^2t^{n-1}}{n-1} - \frac{kt^{1-n}}{A(n-1)}$$

$$\Lambda_1 = \frac{mk^2}{At^n} - 2m\left[Ant^{n-1} - \frac{kt^{1-n}}{A(n-1)} + \frac{An^2t^{n-1}}{n-1}\right](3A^2n^2t^{n-1}(n-1)t^{n-2} + nA^2t^n(n-1)(n-2)t^{n-3}) \qquad 6-7$$

Now from equation(3.3-6)

$$\Lambda_2 = \dot{Q}_2\dot{a} + m\frac{k^2}{a} \qquad 6-8a$$

$$\Lambda_2 = m\frac{k^2}{a} - 2m\left[\ddot{a} + \frac{2a\dot{a}\ddot{a} - \dot{a}k - \dot{a}^3}{a^2}\right]\left(a\dot{a} + \int kdt\right) \qquad 6-8b$$

$$\Lambda_2 = 0$$

$$k = 0$$

$$a\dot{a}\left[\ddot{a} + \frac{2a\dot{a}\ddot{a} - \dot{a}^3}{a^2}\right] = 0$$

$$a\dot{a} = 0 \qquad 6-9$$

$$\ddot{a} + \frac{2a\dot{a}\ddot{a} - \dot{a}^3}{a^2} = 0 \qquad 6-10$$

$$\ddot{a} + \frac{\dot{a}^2}{a} = C \qquad 6-11$$

When $C = 0$,equation (6-11) leads to the same differential equation as equation (6-6). By substituting in equation(6-8b) that $a = At^n$ we find:

$$\Lambda_2 = \frac{mk^2}{At^n} - 2m\left[An(n-1)(n-2)t^{n-3} + \frac{2A^3n^2t^nt^{n-1}(n-1)t^{n-2} - Ant^{n-1}k - A^3n^3t^{3n-3}}{A^2t^{2n}}\right](A^2nt^{n-1}t^n + kt) \qquad 6-12$$

The total cosmological constant is

$$\Lambda_{\text{tot}} = \Lambda_1 + \Lambda_2$$

$$\Lambda_{\text{tot}} = \frac{2m}{A}\{-10A^4n^4t^{3n-4} + 13A^4n^3t^{3n-4} - 2kA^2n^3t^{n-2} + 10kA^2n^2t^{n-2} - 4A^4n^2t^{3n-4} + nk^2t^{-n} - 4kA^2nt^{n-2} + k^2t^{-n}\} \qquad 6-13$$

$$\Lambda = 0 \; when \;\; n = 0 \; or \; \frac{1}{2} \;\; with \; k = 0$$

Cosmological constant vanishes at radiation era.

Now we can calculate the density parameter

$$\Omega_\Lambda = \frac{\Lambda}{3H^2} = \frac{\Lambda t^2}{3n^2}$$

$$\Omega_\Lambda = \frac{2mt^2}{3An^2}\{-10A^4n^4t^{3n-4} + 13A^4n^3t^{3n-4} - 2kA^2n^3t^{n-2} + 10kA^2n^2t^{n-2} - 4A^4n^2t^{3n-4} + nk^2t^{-n} - 4kA^2nt^{n-2} + k^2t^{-n}\} \qquad 6-14$$

$$\rho_\Lambda = \frac{m}{4A\pi G}\{-10A^4n^4t^{3n-4} + 13A^4n^3t^{3n-4} - 2kA^2n^3t^{n-2} + 10kA^2n^2t^{n-2} - 4A^4n^2t^{3n-4} + nk^2t^{-n} - 4kA^2nt^{n-2} + k^2t^{-n}\} \qquad 6-15$$

Now if we study the behavior of $\Lambda$ with respect to n ,k and t we find:

1. $n \to 0$,

$$\Lambda_{\text{n=0}} = \frac{2mk^2}{A} \qquad 6-16$$

$$\rho_\Lambda = \frac{mk^2}{4\pi AG} \qquad 6-17$$

$$\Omega_\Lambda = \frac{2mk^2}{3An^2} t^2 = \frac{2mk^2}{3A} H^{-2} \qquad 6-18$$

From above, when $n \to 0$, cosmological constant and vacuum energy density, remain constants. But the density parameter, vary with time as:

- $t \to \infty, \Omega_\Lambda \to \infty$.this seems to be rejected, because $\Omega_{\text{tot}} = 1$.
- $t \approx n, \Omega_\Lambda = \frac{2mk^2}{3A}$,
- $t \to 0, \Omega_\Lambda \to 0$ .

2. $n = \frac{1}{2}, k \neq 0$

$$\Lambda_{\text{n=0.5}} = \frac{3mk^2}{A\sqrt{t}} + \frac{mkA}{2t^{\frac{2}{3}}}$$

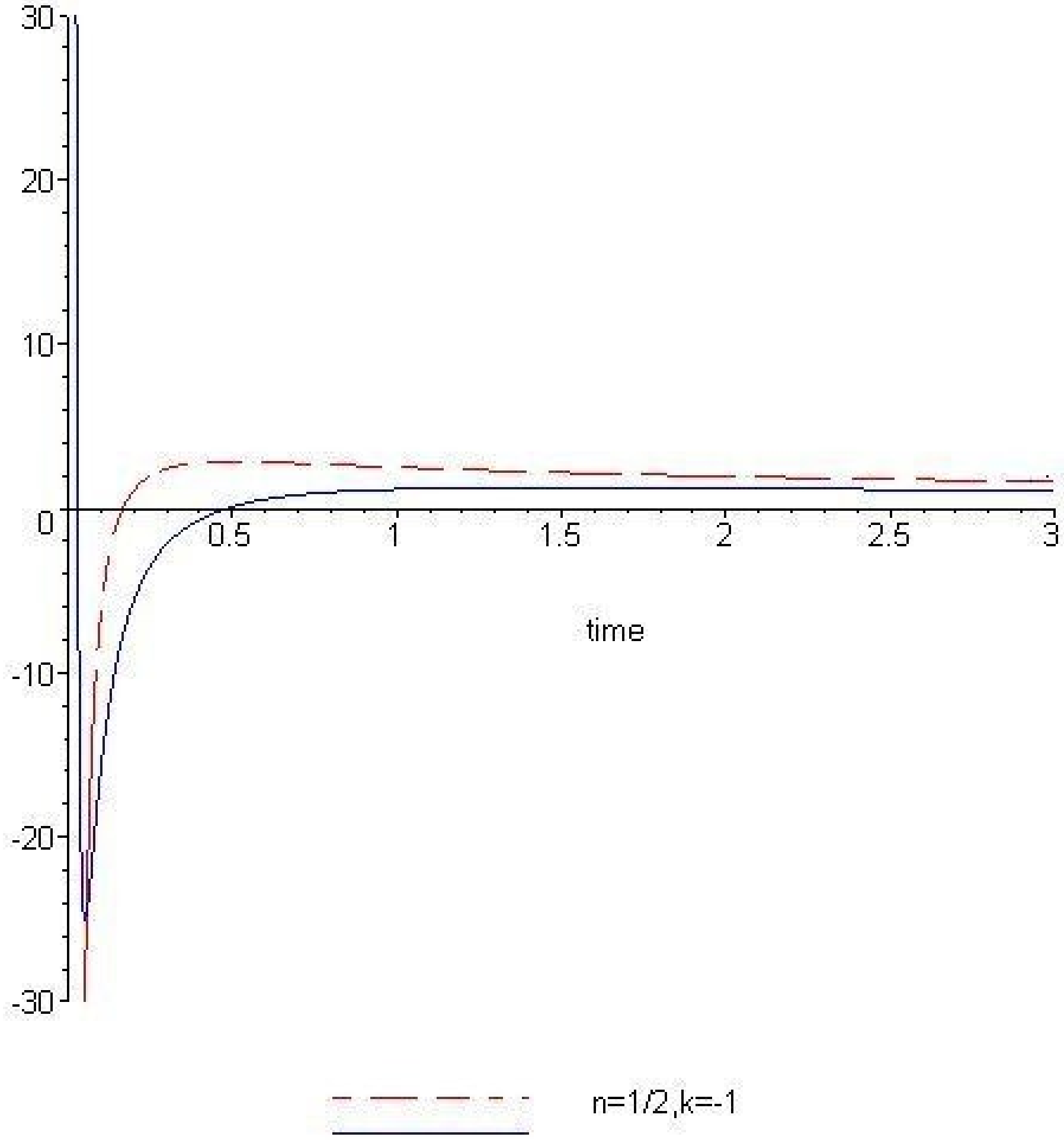

fig$(6-1)$: *shows comparison between cosmological constants* , $n = \frac{1}{2}$ *and* $n = \frac{2}{3}$, $(k = -1)$,

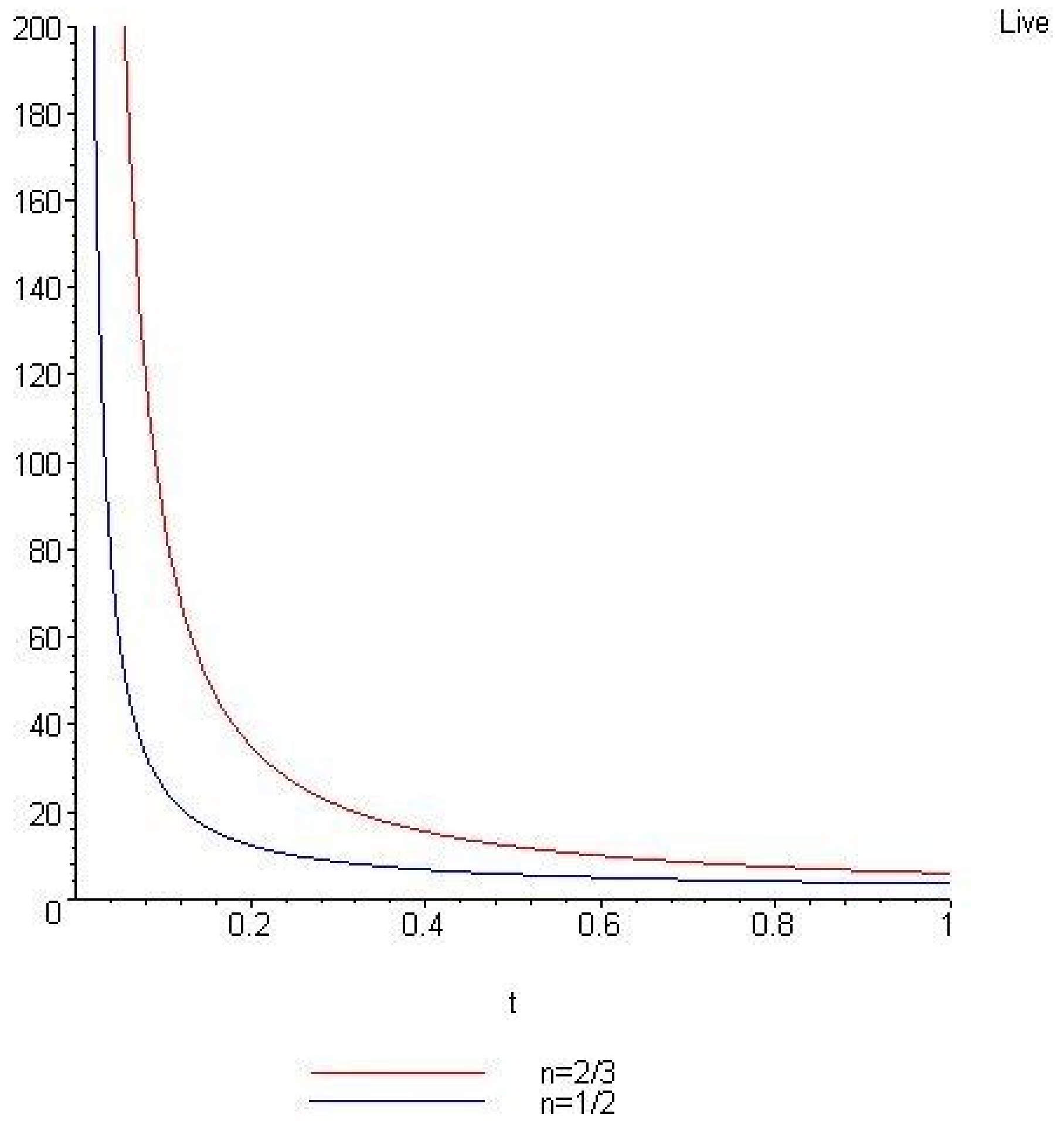

fig$(6-2)$: *shows comparison between cosmological constants* , $n=\frac{1}{2}$ *and* $n=\frac{2}{3}$, $(k=1)$,

$$\Omega_{\Lambda_{n=0.5}}=\frac{4mk^2t^{\frac{3}{2}}}{A}+\frac{2mkAt^{\frac{4}{3}}}{3} \qquad 6-19$$

*7. wave function and quantization:*

Now we can construct the wave function from the above Schrödinger-like equation. Equation ( 2-15) is second order differential equation (H=0):

$$\frac{\hbar^2}{4ma\dot{\gamma}}\frac{\partial^2\psi}{\partial x^2}+\frac{mk^2}{a\dot{\gamma}}\psi=0 \qquad 7-1$$

$$\psi_1=C_1e^{\frac{2imkx}{\hbar}}+C_2e^{\frac{-2imkx}{\hbar}} \qquad 7-2$$

where

$$x=\dot{a}$$

The constructed wave function from second Hamiltonian(3-10)

$$\frac{a\hbar^2}{4m\dot{\alpha}}\frac{\partial^2\psi}{\partial \acute{x}^2}+\frac{mk^2}{a\dot{\alpha}}\psi=0 \qquad 7-3$$

is:

$$\psi_2=D_1e^{\frac{2ikm\acute{x}}{a\hbar}}+D_2e^{\frac{-2ikm\acute{x}}{a\hbar}} \qquad 7-4$$

where

$$\acute{x}=\int \dot{a}da$$

$C_1, C_2, D_1 and\ D_2$ are constants.

At boundaries, (a) is constant , $\psi_1=\psi_2=0$.

Since the wave function depends on k($k=1,-1{,}0$) and $\dot{a}(\dot{a}\geq 0)$.

So k controls the wave function. as the following:

1. $k=0$:

$$\psi_{01}=C_1+C_2 \qquad 7-5$$

and

$$\psi_{02}=D_1+D_2 \qquad 7-6$$

2. $k=1$:

$$\psi_{+1}=C_2e^{\frac{-2imkx}{\hbar}} \qquad 7-7$$

and

$$\psi_{+2}=D_2e^{\frac{-2ikm\acute{x}}{a\hbar}} \qquad 7-8$$

3. $k=-1$:

$$\psi_{-1}=C_1e^{\frac{2imkx}{\hbar}} \qquad 7-9$$

$$\psi_{-2}=D_1e^{\frac{2ikm\acute{x}}{a\hbar}} \qquad 7-10$$

Now if we assume that $a=At^n$

$$\psi(\mathrm{x})=C_1e^{\frac{2imkx}{\hbar}}+D_1e^{\left(\frac{2n}{2n-1}\right)\frac{ikm\,x}{\hbar}}+C_2e^{\frac{-2imkx}{\hbar}}+D_2e^{-\left(\frac{2n}{2n-1}\right)\frac{ikm\,x}{\hbar}} \qquad 7-11$$

$$k=1$$

$$\psi_+=C_2e^{\frac{-2imkx}{\hbar}}+D_2e^{-\left(\frac{2n}{2n-1}\right)\frac{ikm\,x}{\hbar}}$$

$$k=-1$$

$$\psi_-=C_1e^{\frac{2imkx}{\hbar}}+D_1e^{\left(\frac{2n}{2n-1}\right)\frac{ikm\,x}{\hbar}}$$

From boundary conditions when:

- $x=0$ (this is equivalents to $n=0$, we found that:

$$C_1+D_1+C_2+D_2=0 \qquad 7-12$$

- $n=\frac{1}{2}\ and\ k=1$ ,we find that $D_1=0$.if $k=-1$ we find $D_2=0$.

So, the general form of the wave function is:

$$\psi(\mathrm{x}) = \mathrm{C_1 e}^{\frac{2\mathrm{i}|k|\mathrm{mx}}{\hbar}} + \mathrm{C_2 e}^{\frac{-2\mathrm{i}|k|\mathrm{mx}}{\hbar}} - (\mathrm{C_1} + \mathrm{C_2}) e^{-\left(\frac{2n}{2n-1}\right)\frac{i|k|m\,x}{\hbar}} \qquad 7-13$$

We can recognize that from$(7-5, 7-6, 7-12)$ flat universe,$k = 0$, is equivalent to static one $a = constant$ .

1. $n = 1$, $a \propto t$ ,we find the second wave function phase reduced to the first one ,

$$\psi(\mathrm{x}) = \mathrm{C_1 e}^{\frac{2\mathrm{i}|k|\mathrm{mx}}{\hbar}} - \mathrm{C_1 e}^{\frac{-2\mathrm{i}|k|\mathrm{mx}}{\hbar}} = 2i\,\mathrm{C_1} \sin\frac{2|k|m\mathrm{x}}{\hbar} \qquad 7-14$$

2. As we know at radiation era $a \propto t^{\frac{1}{2}}$ , $n = \frac{1}{2}$ ,then, the wave function takes the following form:

$$\psi(\mathrm{x}) = \mathrm{C_1 e}^{\frac{2\mathrm{i}|k|\mathrm{mx}}{\hbar}} + \mathrm{C_2 e}^{\frac{-2\mathrm{i}|k|\mathrm{mx}}{\hbar}} \qquad 7-15$$

We get from 7-14 and 7-15 $C_1 = -C_2 \,.D_2 = 0$

Using 7-14 to find out $\mathrm{C_1}$ by normalizing the wave function, $\int \psi\psi^* \mathrm{dx} = 1$

$$\int_0^A \left|2i\,\mathrm{C_1} \sin\frac{2|k|m\mathrm{x}}{\hbar}\right|^2 \mathrm{dx} = 1$$

$$C_1 = \sqrt{\frac{b}{\sin 2Ab - 2Ab}} \qquad 7-16$$

Where : $b = \frac{2|k|m}{\hbar}$ .

According to this limit $-\infty \leq \frac{b}{\sin 2Ab - 2Ab} \leq -1$ and since $A \geq 0$ , $C_1$ is complex number .

We can write 7-14 as:

$$\psi(\dot{a}) = 2\sqrt{\frac{b}{2Ab - \sin 2Ab}} \sin b\dot{a} \qquad 7-17$$

Also, we found that

$$0 \leq n \leq 1 \qquad 7-18$$

According to 5-5 and 7-18 , at early universe vacuum dose not dominate the universe. So the limit of $\omega$ is reduced to:

$$-\frac{1}{3} \leq \omega \leq \frac{1}{n} \qquad 7-19$$

We can find the probability, which we can call it density parameter $\Omega$, to our universe, from 7-17

$$P \equiv \Omega = \frac{4b}{2Ab - \sin 2Ab} \sin^2(b\dot{a}) \qquad 7-20a$$

$$P \equiv \Omega = \frac{8|k|m}{4A|k|m - \hbar \sin\frac{4A|k|m}{\hbar}} \sin^2\left(\frac{2|k|m}{\hbar}\dot{a}\right) \qquad 7-20b$$

When $2Ab \gg \sin 2Ab$ , we find that $\Omega = \frac{2}{A}\sin^2(b\dot{a})$.

If we take the probability in flat and unaccelerated universe(appindixI-3),

$$\lim_{k=0, \dot{a}=A} \Omega = \frac{3}{A} \qquad 7-21$$

Now ,if we take the limit of the probability in flat and accelerated universe,

$$\lim_{k=0} \Omega = \frac{3\dot{a}^2}{A^3} \qquad 7-22$$

Using Friedmann equation ,where $k = 0$ we find:

$$\lim_{k=0} \Omega = \frac{8\pi G\rho a^2}{A^3} \qquad 7-23$$

Maximum probability, $\Omega = 1$ , occurs when, $k \neq 0$, and:

$$\dot{a}_c = \frac{\hbar}{2m}\sin^{-1}\sqrt{\frac{4Am - \hbar\sin\frac{4Am}{\hbar}}{8m}} \qquad 7-24$$

Where $a \propto t^n$ .

$$t_c = \left[\frac{\hbar}{2Anm}\sin^{-1}\sqrt{\frac{4Am - \hbar\sin\frac{4Am}{\hbar}}{8m}}\right]^{\frac{1}{n-1}} \qquad 7-25$$

$$a_c = A\left[\frac{\hbar}{2Anm}\sin^{-1}\sqrt{\frac{4Am - \hbar\sin\frac{4Am}{\hbar}}{8m}}\right]^{\frac{n}{n-1}} \qquad 7-26$$

$$H_c = n^{\left(\frac{n}{n-1}\right)}\left[\frac{\hbar}{2Am}\sin^{-1}\sqrt{\frac{4Am - \hbar\sin\frac{4Am}{\hbar}}{8m}}\right]^{\frac{1}{1-n}} \qquad 7-27$$

$$\lim_{n\to 0} a_c = A \qquad 7-28$$

$$\lim_{n\to 1} a_c = \infty \qquad 7-29$$

$$\lim_{n\to 0} H_c = \frac{\dot{a}_c}{A} \qquad 7-30$$

The subscript( $c$) refers to critical value ,which occurs at maximum probability.

From 7-20, $\Omega = \frac{8\pi G\rho}{3H^2}$ , we find:

$$\rho = \frac{3|k|m}{\pi G}\,\frac{H^2}{\left[4A|k|m - \hbar\sin\frac{4A|k|m}{\hbar}\right]}\,\sin^2\left(\frac{2|k|m\dot{a}}{\hbar}\right) \qquad 7-31a$$

Where $k \neq 0$, we find:

$$\rho = \frac{3m}{\pi G}\,\frac{H^2}{\left[4Am - \hbar\sin\frac{4Am}{\hbar}\right]}\,\sin^2\left(\frac{2m\dot{a}}{\hbar}\right) \qquad 7-31b$$

$$\rho_c = \frac{3}{8\pi G}\,\frac{n^{\left(\frac{2n}{n-1}\right)}}{\left[4Am - \hbar\sin\frac{4Am}{\hbar}\right]}\left[\frac{\hbar}{2Am}\sin^{-1}\sqrt{\frac{4Am - \hbar\sin\frac{4Am}{\hbar}}{8m}}\right]^{\frac{2}{1-n}}\left[4Am - \hbar\, sin\frac{4Am}{\hbar}\right] \quad 7-32$$

7-31,must satisfy the condition:

$$A \times m > \frac{\hbar}{4} \qquad 7-33$$

If we consider, that,$\dot{a} \to 0$ ; 7-31 reads:

$$\rho_{|\dot{a}\to 0} = \frac{12m^3k^2}{\pi\hbar^2 G}\,\frac{H^4a^2}{\left[4Am - \hbar\sin\frac{4Am}{\hbar}\right]} \qquad 7-34$$

Now we define the continuity equation:

$$\frac{\partial P}{\partial t} + \bar{\nabla}.\bar{J} = 0 \qquad 7-35$$

$$\text{J(current density)} = \frac{i\hbar}{2m}\left[\Psi\frac{\partial\Psi^*}{\partial\dot{a}} - \Psi^*\frac{\partial\Psi}{\partial\dot{a}}\right] = 0 \qquad 7-36$$

$$\frac{\partial P}{\partial t} = \frac{32\,k^2m^2n(n-1)t^{(n-1)}}{\left[4A|k|m - \hbar\, sin\frac{4A|k|m}{\hbar}\right]\hbar t}\left[\cos\frac{2An|k|mt^{(n-1)}}{\hbar}\sin\frac{2An|k|mt^{(n-1)}}{\hbar}\right] = 0$$

Now we can solve for (t),(n).(appindixI-4);

$$t(n) = \exp\left(\frac{\ln\frac{\pm\pi\hbar}{4A|k|nm}}{n-1}\right); t = 0 \qquad 7-37$$

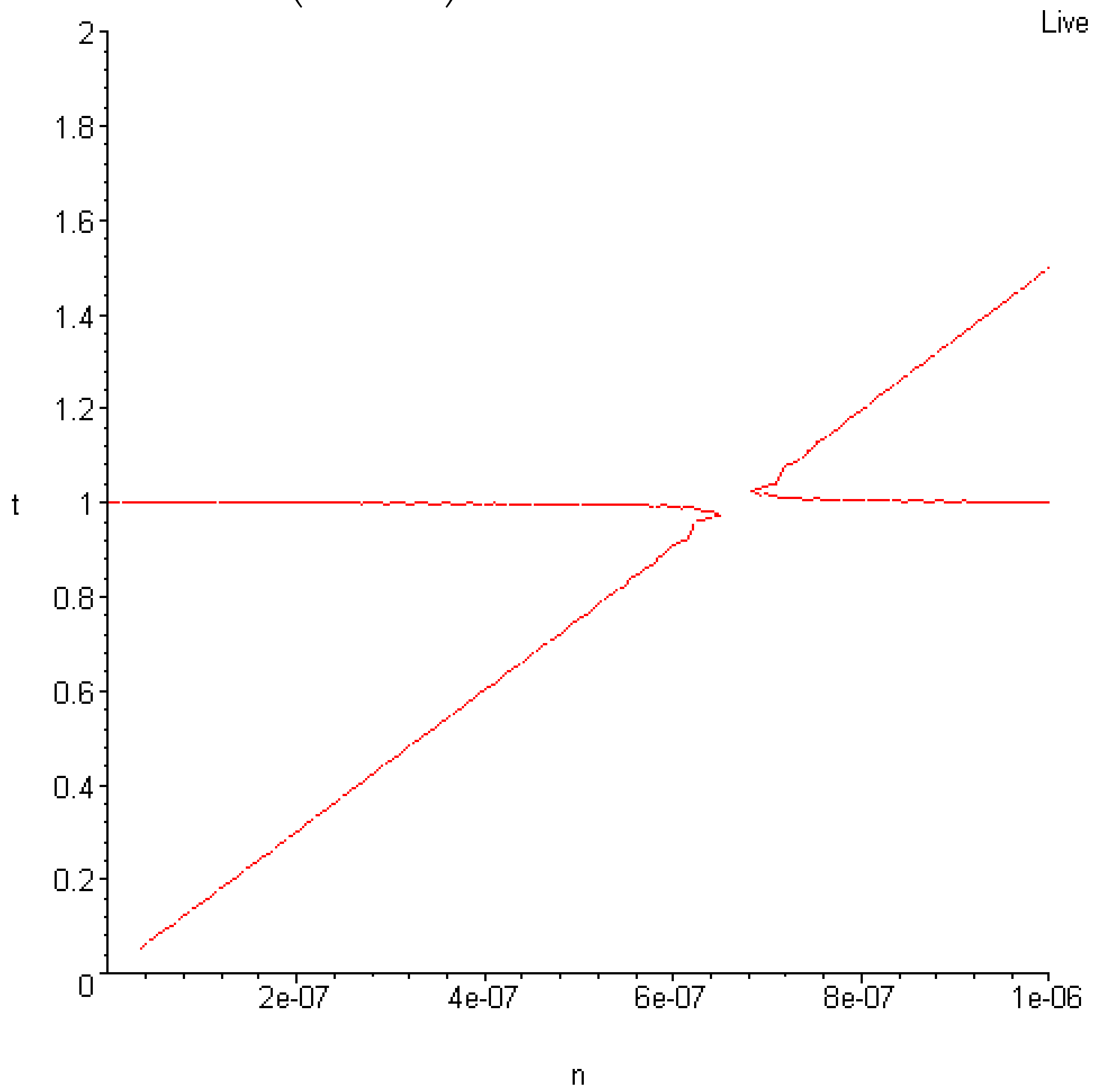

fig$(7-1)$: *shows variation of* $t-n$, *when* $t \to 1$, *we find* $n$ *is indefinit.*

$$n(t) = \frac{\mathrm{LambertW}\left(\frac{\pm\pi\hbar}{4A|k|m} t \ln t\right)}{\ln t}; n = 0; n = 1 \qquad 7-38$$

7-38 gives inflation at (t=1).

Age of our universe is $10^9\ y$,(appendix I),$n = \frac{2}{3}$, from 7-37 we find:

$$\frac{4Am|k|}{\pi\hbar} = 1.059505956 \times 10^6\ ; m \approx 2.65\ \times 10^{-29}\ (j.s^2)$$

Now we can estimate radiation dominated age, with $n = \frac{1}{2}$;

$$t_{rad}\ \approx 10^5 y$$

Now we find the scale factor, from 7-37 and 7-38(appendix I- 6) as:

$$a(n) = A\left[\exp\left(\frac{ln\frac{\pm\pi\hbar}{4A|k|nm}}{n-1}\right)\right]^n \qquad 7-39$$

And $\dot{a}$ as:

$$\dot{a}(n) = An\left[\exp\left(\frac{ln\frac{\pm\pi\hbar}{4A|k|nm}}{n-1}\right)\right]^{n-1} \qquad 7-40$$

And acceleration is given by(appendix I- 7):

$$\ddot{a}(n) = An(n-1)\left[\exp\left(\frac{ln\frac{\pm\pi\hbar}{4A|k|nm}}{n-1}\right)\right]^{n-2} \qquad 7-41$$

$$H(n) = n\left[\exp\left(\frac{ln\frac{\pm\pi\hbar}{4A|k|nm}}{n-1}\right)\right]^{-1} \qquad 7-42$$

appendix I-8 to calculate mass density and Hubble constant .
From 7-24 and( appendix I-6) ,we can estimate the value of

$$\sin^{-1}\sqrt{\frac{4Am - \hbar\sin\frac{4Am}{\hbar}}{8m}} = \pi/2;\ or\ \frac{4Am - \hbar\, sin\frac{4Am}{\hbar}}{8m} = 1 \qquad 7-43$$

***8. Complex solution:***

Now we solved 7-37 in context of the term $\frac{\pi\hbar}{4A|k|m}$, according to (appendix I-4) we find:

$$\frac{\pi\hbar}{4A|k|m} = -4.4\times 10^{-7} \pm 7.6\times 10^{-7}i \qquad 8-1$$

If we assume that, the constant ,A, is a complex, using 8-1,we find:

$$A = (1 + 1.7321i)\pi, \qquad A^* = (1 - 1.7321i)\pi \qquad 8-2$$

$$\frac{4Am - \hbar\, sin\frac{4Am}{\hbar}}{8m} = -.1276 + .2265i \qquad 8-3$$

By substituting 8-1 and 8-2 in 7-39( appendix I-9) ,7-40( appendix I-10),7-41( appendix I-11)and 7-42( appendix I-12).
From appendix I-11, our today universe acceleration ($n = \frac{2}{3}$) can be assigned to the imaginary part of $\ddot{a}$. Also we can find the density, by substituting 8-3 in 7-31, ( appendix I-13).

If we treat the cosmic back ground radiation as black body radiation, and according to Stefan`s law of radiation:

$$\sigma T^4 \propto E$$

$$T = \sqrt[4]{\left[\frac{4}{3\sigma}\pi A^3\left(\frac{a_0}{a_r}\right)^3 f\rho_r\left(\frac{1}{(1+z)}\right)^2\right]} \qquad 8-4$$

Where $f$ is the ratio of redshifted radiation density.

$$0 \le f \le 1$$

8-3 represents maximum(limit) CBR temperature when $f = 1$
If we substitute the values from which we found ,from our work, we get:

$$f = 1$$
$$T = 5.3 - 3.1i\ k$$

When $f = 0.07$

$$T = 2.726 - 1.58i\ k$$

Now if we choose the scale factor with variable A as:

$$a = A^n t^n \qquad 8-5$$

Now we can rewrite our equations as:

$$P \equiv \Omega = \frac{8|k|m}{4A^n|k|m - \hbar \sin\frac{4A^n|k|m}{\hbar}} \sin^2\left(\frac{2|k|m}{\hbar}\dot{a}\right) \qquad 8-6$$

$$\rho = \frac{3|k|m}{\pi G} \frac{H^2}{\left[4A^n|k|m - \hbar \sin\frac{4A^n|k|m}{\hbar}\right]} \sin^2\left(\frac{2|k|m\dot{a}}{\hbar}\right) \qquad 8-7$$

$$t(n) = \exp\left(\frac{\ln\frac{\pm\pi\hbar}{4|k|nm} - n\ln A}{n-1}\right); t = 0 \qquad 8-9$$

$$n(t) = \frac{\mathrm{LambertW}\left(\frac{\pm\pi\hbar}{4A|k|m} t \ln At\right)}{\ln At}; n = 0; n = 1 \qquad 8-10$$

$$T = (1.89654434 - 1.8965443381i)\ K$$

With absolute value,$T = 2.682118726\ K$

When the age of universe is:

$$t = 13.75\ Gy$$

Which is agree with recent observations.

## *9. Results and Discussion:*

We have seen that the canonical quantization formalism adopted in this work has brought interesting results. We have solved the Wheeler-De Witt equation and obtained the wave function of the universe. The wave function vanishes when $\dot{a} \to 0$ . We found that density parameter is related to probability. Moreover, we have found that the scale factor is imaginary. From Stefan's law we have found the cosmic microwave background radiation (CMBR) temperature.

Considering the model of $\ddot{a}$ it is important to note that in case of $a \to \infty , \dot{a} \ll 1$ , i.e. when the universe radius is large, the expansion rate is slow, equation (2-26) which is the expression for the density reduces to that of GR, the density of universe decreases with time as radius increases in conformity with observations. It also indicates that the pressure, P, vanishes, equation 2.28, which also agrees with GR and observations at matter era.

Thus, this case represents matter era in GR. When the radius is assumed to be very small with high expansion rate i.e. $a \to 0 , \dot{a} \to \infty$, as happens at early universe, the pressure becomes negative ,equation 2.32. This also agrees with inflation assumption.

The model for $\dot{a}^2$ gives also some interesting results, it shows that $\rho \propto a^{-3}$ and $p = 0$ which can suitably describing matter era only.

The boundary term $\sigma$ for this model coincides with the one emerging from that of the model $\ddot{a}$, as shown by equation 2.9.

The two cosmological constant $\Lambda_1$ and $\Lambda_2$ for the model $\ddot{a}$ and $\dot{a}^2$ show also some interesting properties. They both vanishes at radiation era, when $a = At^{\frac{1}{2}}$.
As seen in equations (6-2), (6-7), (6.9) and (6-13), equation (6-19) shows that the total cosmological constant decreases as time increases. This agrees with inflation models at the early universe.
The wavefunction $\psi_1$ and $\psi_2$ of the universe for both models are also obtained in section [7]. The complex form of the solution in (7-13) shows the existance of quantum effects. The complex nature of cosmic scale factor shown in (8-2) confirms the existance of quantum effects. The CBR temperature is found to agree with the experimental range , equation 8-5.

*APPENDIX*

1. This program (in maple code) to calculate the total cosmological constant,here we referred to $\Lambda_{tot}$ as (ty).

```
> ty:=proc(t,n,k);
> localresult;
> result:
> 2*m*k^2/A/(t^n)-2*m*(A*n*t^(n-1)-t^(-n+1)/A/(n-1)*k+t^(n-1)*A/(n-1)*n^2)*(3*A^2*n^2*t^(n-1)*(n-1)*t^(n-2)+A^2*t^n*n*(n-1)*(n-2)*t^(n-3))-2*m*(A*n*(n-1)*(n-2)*t^(n-3)+(2*A^3*t^n*n^2*t^(n-1)*(n-1)*t^(n-2)-A*n*t^(n-1)*k-A^3*n^3*(t^(n-1))^3)/A^2/(t^n)^2)*(A^2*t^n*n*t^(n-1)+k*t);
> end;
```

$ty := \textbf{proc}(t, n, k)$

$localresult;$

$result;$

$$2\times m\times k\hat{}2/(A\times t\hat{}n) - 2\times m\times (A\times n\times t\hat{}(n-1) - t\hat{}(-n+1)\times k/(A\times(n-1)) + t\hat{}(n-1)\times A\times n\hat{}2/(n-1))\times (3\times A\hat{}2\times n\hat{}2\times t\hat{}(n-1)\times(n-1)\times t\hat{}(n-2) + A\hat{}2\times t\hat{}n\times n\times(n-1)\times(n-2)\times t\hat{}(n-3)) - 2\times m\times( A\times n\times(n-1)\times(n-2)\times t\hat{}(n-3) + ( 2\times A\hat{}3\times t\hat{}n\times n\hat{}2\times t\hat{}(n-1)\times(n-1)\times t\hat{}(n-2) - A\times n\times t\hat{}(n-1)\times k - A\hat{}3\times n\hat{}3\times(t\hat{}(n-1))\hat{}3)/(A\hat{}2\times(t\hat{}n)\hat{}2))\times(A\hat{}2\times t\hat{}n\times n\times t\hat{}(n-1) + k\times t)$$

**end proc**

```
> ty(t,n,k);
```

$$\frac{2\,m\,k^2}{A\,t^n} - 2\,m\left(A\,n\,t^{(n-1)} - \frac{t^{(-n+1)}\,k}{A\,(n-1)} + \frac{t^{(n-1)}\,A\,n^2}{n-1}\right)$$

$$(3\,A^2\,n^2\,t^{(n-1)}\,(n-1)\,t^{(n-2)} + A^2\,t^n\,n\,(n-1)\,(n-2)\,t^{(n-3)}) - 2\,m\Bigg($$

$$A\,n\,(n-1)\,(n-2)\,t^{(n-3)}$$

$$+\frac{2\,A^3\,t^n\,n^2\,t^{(n-1)}\,(n-1)\,t^{(n-2)} - A\,n\,t^{(n-1)}\,k - A^3\,n^3\,(t^{(n-1)})^3}{A^2\,(t^n)^2}\Bigg)$$

$$(A^2\,t^n\,n\,t^{(n-1)} + k\,t)$$

```
> R0 := simplify(2*m*k^2/A/(t^n)-2*m*(A*n*t^(n-1)-t^(-
n+1)/A/(n-1)*k+t^(n-1)*A/(n-1)*n^2)*(3*A^2*n^2*t^(n-
1)*(n-1)*t^(n-2)+A^2*t^n*n*(n-1)*(n-2)*t^(n-3))-
2*m*(A*n*(n-1)*(n-2)*t^(n-3)+(2*A^3*t^n*n^2*t^(n-1)*(n-
1)*t^(n-2)-A*n*t^(n-1)*k-A^3*n^3*(t^(n-
1))^3)/A^2/(t^n)^2)*(A^2*t^n*n*t^(n-1)+k*t));
```

$$R0 := -2\,(10\,A^4\,n^4\,t^{(3\,n-4)} - 13\,A^4\,n^3\,t^{(3\,n-4)} + 2\,A^2\,n^3\,t^{(n-2)}\,k - 10\,A^2\,n^2\,t^{(n-2)}\,k + 4\,A^4\,n^2\,t^{(3\,n-4)} - t^{(-n)}\,n\,k^2 + 4\,A^2\,n\,t^{(n-2)}\,k - t^{(-n)}\,k^2)\,m/A$$

```
> R1 := combine(R0,power);
```

$$R1 := -2\,(10\,A^4\,n^4\,t^{(3\,n-4)} - 13\,A^4\,n^3\,t^{(3\,n-4)} + 2\,A^2\,n^3\,t^{(n-2)}\,k - 10\,A^2\,n^2\,t^{(n-2)}\,k + 4\,A^4\,n^2\,t^{(3\,n-4)} - t^{(-n)}\,n\,k^2 + 4\,A^2\,n\,t^{(n-2)}\,k - t^{(-n)}\,k^2)\,m/A$$

```
> ty(t,0,0);
```

$$0$$

```
> ty(t,.5,0);
```

$$0.$$

```
> ty(t,1/2,-1);
```

$$\frac{2\,m}{A\,\sqrt{t}} - 2\,m\left(\frac{3\,A}{8\,t^{(5/2)}} + \frac{-\frac{3\,A^3}{8\,t^{(3/2)}} + \frac{A}{2\,\sqrt{t}}}{A^2\,t}\right)\left(\frac{A^2}{2} - t\right)$$

```
> R18 := simplify(2*m/A/t^(1/2)-2*m*(3/8*A/t^(5/2)+(-
3/8*A^3/t^(3/2)+1/2*A/t^(1/2))/A^2/t)*(1/2*A^2-t));
```

$$R18 := \frac{m\,(6\,t - A^2)}{2\,t^{(3/2)}\,A}$$

```
> R19 := expand(R18);
```

$$R19 := \frac{3\,m}{A\,\sqrt{t}} - \frac{m\,A}{2\,t^{(3/2)}}$$

```
> ty(t,1/2,1);
```

$$\frac{2\,m}{A\,\sqrt{t}} - 2\,m\left(\frac{3\,A}{8\,t^{(5/2)}} + \frac{-\frac{3\,A^3}{8\,t^{(3/2)}} - \frac{A}{2\,\sqrt{t}}}{A^2\,t}\right)\left(\frac{A^2}{2} + t\right)$$

```
> R20 := simplify(2*m/A/t^(1/2)-2*m*(3/8*A/t^(5/2)+(-3/8*A^3/t^(3/2)-1/2*A/t^(1/2))/A^2/t)*(1/2*A^2+t));
```

$$R20 := \frac{m\,(6\,t + A^2)}{2\,t^{(3/2)}\,A}$$

```
> R21 := expand(R20);
```

$$R21 := \frac{3\,m}{A\,\sqrt{t}} + \frac{m\,A}{2\,t^{(3/2)}}$$

```
> ty(t,2/3,-1);
```

$$\frac{2\,m}{A\,t^{(2/3)}} + \frac{8\,m\left(-\frac{2\,A}{3\,t^{(1/3)}} - \frac{3\,t^{(1/3)}}{A}\right)A^2}{27\,t^{(5/3)}} - 2\,m\left(\frac{8\,A}{27\,t^{(7/3)}} + \frac{-\frac{16\,A^3}{27\,t} + \frac{2\,A}{3\,t^{(1/3)}}}{A^2\,t^{(4/3)}}\right)\left(\frac{2\,A^2\,t^{(1/3)}}{3} - t\right)$$

```
> R22 := simplify(2*m/A/t^(2/3)+8/27*m*(-2/3*A/t^(1/3)-3*t^(1/3)/A)*A^2/t^(5/3)-2*m*(8/27*A/t^(7/3)+(-16/27*A^3/t+2/3*A/t^(1/3))/A^2/t^(4/3))*(2/3*A^2*t^(1/3)-t));
```

$$R22 := -\frac{2\,m\,(-135\,t^{(5/3)} - 8\,A^4\,t^{(1/3)} + 96\,A^2\,t)}{81\,t^{(7/3)}\,A}$$

```
> R23 := expand(R22);
```

$$R23 := \frac{10\,m}{3\,A\,t^{(2/3)}} + \frac{16\,m\,A^3}{81\,t^2} - \frac{64\,m\,A}{27\,t^{(4/3)}}$$

```
> ty(t,2/3,1);
```

$$\frac{2\,m}{A\,t^{(2/3)}} + \frac{8\,m\left(-\frac{2\,A}{3\,t^{(1/3)}} + \frac{3\,t^{(1/3)}}{A}\right)A^2}{27\,t^{(5/3)}} - 2\,m\left(\frac{8\,A}{27\,t^{(7/3)}} + \frac{-\frac{16\,A^3}{27\,t} - \frac{2\,A}{3\,t^{(1/3)}}}{A^2\,t^{(4/3)}}\right)\left(\frac{2\,A^2\,t^{(1/3)}}{3} + t\right)$$

```
> R24 := simplify(2*m/A/t^(2/3)+8/27*m*(-2/3*A/t^(1/3)+3*t^(1/3)/A)*A^2/t^(5/3)-2*m*(8/27*A/t^(7/3)+(-16/27*A^3/t-2/3*A/t^(1/3))/A^2/t^(4/3))*(2/3*A^2*t^(1/3)+t));
```

$$R24 := \frac{2\,m\,(135\,t^{(5/3)} + 8\,A^4\,t^{(1/3)} + 96\,A^2\,t)}{81\,t^{(7/3)}\,A}$$

```
> R25 := expand(R24);
```

$$R25 := \frac{10\,m}{3\,A\,t^{(2/3)}} + \frac{16\,m\,A^3}{81\,t^2} + \frac{64\,m\,A}{27\,t^{(4/3)}}$$

```
> ty(t,n,0);
```

$$-2\,m\left(A\,n\,t^{(n-1)}+\frac{t^{(n-1)}\,A\,n^2}{n-1}\right)(3\,A^2\,n^2\,t^{(n-1)}\,(n-1)\,t^{(n-2)}+A^2\,t^n\,n\,(n-1)\,(n-2)\,t^{(n-3)})-2\,m\left(A\,n\,(n-1)\,(n-2)\,t^{(n-3)}+\frac{2\,A^3\,t^n\,n^2\,t^{(n-1)}\,(n-1)\,t^{(n-2)}-A^3\,n^3\,(t^{(n-1)})^3}{A^2\,(t^n)^2}\right)A^2\,t^n\,n\,t^{(n-1)}$$

```
> ty(t,1/2,k);
```

$$\frac{2\,m\,k^2}{A\,\sqrt{t}}-2\,m\left(\frac{3\,A}{8\,t^{(5/2)}}+\frac{-\frac{3\,A^3}{8\,t^{(3/2)}}-\frac{A\,k}{2\,\sqrt{t}}}{A^2\,t}\right)\left(\frac{A^2}{2}+k\,t\right)$$

```
> R0 := simplify(2*m*k^2/A/t^(1/2)-2*m*(3/8*A/t^(5/2)+(-3/8*A^3/t^(3/2)-1/2*A/t^(1/2)*k)/A^2/t)*(1/2*A^2+k*t));
```

$$R0 := \frac{m\,k\,(6\,k\,t+A^2)}{2\,t^{(3/2)}\,A}$$

```
> R1 := expand(R0);
```

$$R1 := \frac{3\,m\,k^2}{A\,\sqrt{t}}+\frac{m\,k\,A}{2\,t^{(3/2)}}$$

```
> limit(R1,t=infinity);
```

$$0$$

2. This is maple code to calculate $\Lambda$, $\Omega_\Lambda$ and $\rho_\Lambda$

```
> lmabda=m*k^2/A/(t^n)-2*m*(A*n*t^(n-1)-t^(-n+1)/A/(n-1)*k+t^(n-1)*A/(n-1)*n^2)*(3*A^2*n^2*t^(n-1)*(n-1)*t^(n-2)+A^2*t^n*n*(n-1)*(n-2)*t^(n-3))+m*k^2/A/(t^n)-2*m*(A*n*(n-1)*(n-2)*t^(n-3)+(2*A^3*t^n*n^2*t^(n-1)*(n-1)*t^(n-2)-A*n*t^(n-1)*k-A^3*n^3*(t^(n-1))^3)/A^2/(t^n)^2)*(A^2*t^n*n*t^(n-1)+k*t);
```

$$lmabda=\frac{2\,m\,k^2}{A\,t^n}-2\,m\left(A\,n\,t^{(n-1)}-\frac{t^{(-n+1)}\,k}{A\,(n-1)}+\frac{t^{(n-1)}\,A\,n^2}{n-1}\right)(3\,A^2\,n^2\,t^{(n-1)}\,(n-1)\,t^{(n-2)}+A^2\,t^n\,n\,(n-1)\,(n-2)\,t^{(n-3)})-2\,m\left(A\,n\,(n-1)\,(n-2)\,t^{(n-3)}+\frac{2\,A^3\,t^n\,n^2\,t^{(n-1)}\,(n-1)\,t^{(n-2)}-A\,n\,t^{(n-1)}\,k-A^3\,n^3\,(t^{(n-1)})^3}{A^2\,(t^n)^2}\right)(A^2\,t^n\,n\,t^{(n-1)}+k\,t)$$

```
> R16 := rhs(lambda = 2*m*k^2/A/(t^n)-2*m*(A*n*t^(n-1)-t^(-n+1)/A/(n-1)*k+t^(n-1)*A/(n-1)*n^2)*(3*A^2*n^2*t^(n-
```

```
1)*(n-1)*t^(n-2)+A^2*t^n*n*(n-1)*(n-2)*t^(n-3))-
2*m*(A*n*(n-1)*(n-2)*t^(n-3)+(2*A^3*t^n*n^2*t^(n-1)*(n-
1)*t^(n-2)-A*n*t^(n-1)*k-A^3*n^3*(t^(n-
1))^3)/A^2/(t^n)^2)*(A^2*t^n*n*t^(n-1)+k*t));
```

$$R16 := \frac{2\,m\,k^2}{A\,t^n} - 2\,m\left(A\,n\,t^{(n-1)} - \frac{t^{(-n+1)}\,k}{A\,(n-1)} + \frac{t^{(n-1)}\,A\,n^2}{n-1}\right) (3\,A^2\,n^2\,t^{(n-1)}\,(n-1)\,t^{(n-2)} + A^2\,t^n\,n\,(n-1)\,(n-2)\,t^{(n-3)}) - 2\,m\left(A\,n\,(n-1)\,(n-2)\,t^{(n-3)} + \frac{2\,A^3\,t^n\,n^2\,t^{(n-1)}\,(n-1)\,t^{(n-2)} - A\,n\,t^{(n-1)}\,k - A^3\,n^3\,(t^{(n-1)})^3}{A^2\,(t^n)^2}\right)(A^2\,t^n\,n\,t^{(n-1)} + k\,t)$$

```
> R17 := simplify(R16);
```

$$R17 := -2\,(10\,A^4\,n^4\,t^{(3\,n-4)} - 13\,A^4\,n^3\,t^{(3\,n-4)} + 2\,A^2\,n^3\,t^{(n-2)}\,k - 10\,A^2\,n^2\,t^{(n-2)}\,k + 4\,A^4\,n^2\,t^{(3\,n-4)} - t^{(-n)}\,n\,k^2 + 4\,A^2\,n\,t^{(n-2)}\,k - t^{(-n)}\,k^2)\,m/A$$

```
> R5 := expand(R17);
```

$$R5 := -\frac{20\,m\,A^3\,n^4\,(t^n)^3}{t^4} + \frac{26\,m\,A^3\,n^3\,(t^n)^3}{t^4} - \frac{4\,m\,A\,n^3\,t^n\,k}{t^2} + \frac{20\,m\,A\,n^2\,t^n\,k}{t^2} - \frac{8\,m\,A^3\,n^2\,(t^n)^3}{t^4} + \frac{2\,m\,n\,k^2}{A\,t^n} - \frac{8\,m\,A\,n\,t^n\,k}{t^2} + \frac{2\,m\,k^2}{A\,t^n}$$

```
> R6 := collect(R5,k);
```

$$R6 := \left(\frac{2\,m\,n}{A\,t^n} + \frac{2\,m}{A\,t^n}\right)k^2 + \left(-\frac{4\,m\,A\,n^3\,t^n}{t^2} + \frac{20\,m\,A\,n^2\,t^n}{t^2} - \frac{8\,m\,A\,n\,t^n}{t^2}\right)k - \frac{20\,m\,A^3\,n^4\,(t^n)^3}{t^4} + \frac{26\,m\,A^3\,n^3\,(t^n)^3}{t^4} - \frac{8\,m\,A^3\,n^2\,(t^n)^3}{t^4}$$

This is $\Omega_\Lambda$

```
> omega:=R17*t^2/(3*n^2);
```

$$\omega := -2\,(10\,A^4\,n^4\,t^{(3\,n-4)} - 13\,A^4\,n^3\,t^{(3\,n-4)} + 2\,A^2\,n^3\,t^{(n-2)}\,k - 10\,A^2\,n^2\,t^{(n-2)}\,k + 4\,A^4\,n^2\,t^{(3\,n-4)} - t^{(-n)}\,n\,k^2 + 4\,A^2\,n\,t^{(n-2)}\,k - t^{(-n)}\,k^2)\,m\,t^2 / (3\,A\,n^2)$$

```
> R7 := collect(omega,k);
```

$$R7 := -\frac{2\,(-t^{(-n)}\,n - t^{(-n)})\,m\,t^2\,k^2}{3\,A\,n^2} - \frac{2\,(2\,A^2\,n^3\,t^{(n-2)} - 10\,A^2\,n^2\,t^{(n-2)} + 4\,A^2\,n\,t^{(n-2)})\,m\,t^2\,k}{3\,A\,n^2} - \frac{2\,(10\,A^4\,n^4\,t^{(3\,n-4)} - 13\,A^4\,n^3\,t^{(3\,n-4)} + 4\,A^4\,n^2\,t^{(3\,n-4)})\,m\,t^2}{3\,A\,n^2}$$

Here we referred to $\rho_\Lambda$ as (q):

```
> q:=R17/(8*pi*G);
```

$$q := -(10A^4 n^4 t^{(3n-4)} - 13A^4 n^3 t^{(3n-4)} + 2A^2 n^3 t^{(n-2)} k - 10A^2 n^2 t^{(n-2)} k + 4A^4 n^2 t^{(3n-4)} - t^{(-n)} n k^2 + 4A^2 n t^{(n-2)} k - t^{(-n)} k^2)\, m/(4A\pi G)$$

```
> limit(R17,n=0);
```

$$\frac{2k^2 m}{A}$$

3. This program to calculate the wave function $\psi$

```
> wave:=proc(b,A,x);
> localresult;
> result:psi(x)=2*sqrt(b/(2*A*b-sin(2*A*b)))*sin(b*x);
> end;
```

*wave* := **proc**$(b, A, x)$

$\quad$ *localresult*; *result*; $\psi(x) = 2\times\text{sqrt}(b/(2\times A\times b - \sin(2\times A\times b)))\times\sin(b\times x)$

**end proc**

```
> wave(b,A,x);
```

$$\psi(x) = 2\sqrt{\frac{b}{2Ab - \sin(2Ab)}}\,\sin(bx)$$

```
> wave(2*abs(k)*m/h,A,x);
```

$$\psi(x) = 2\sqrt{2}\sqrt{\frac{|k|m}{h\left(\frac{4A|k|m}{h} - \sin\left(\frac{4A|k|m}{h}\right)\right)}}\,\sin\left(\frac{2|k|mx}{h}\right)$$

```
> R0 := rhs(psi(x) =
2*2^(1/2)*(abs(k)*m/h/(4*A*abs(k)*m/h-
sin(4*A*abs(k)*m/h)))^(1/2)*sin(2*abs(k)*m/h*x));
```

$$R0 := 2\sqrt{2}\sqrt{\frac{|k|m}{h\left(\frac{4A|k|m}{h} - \sin\left(\frac{4A|k|m}{h}\right)\right)}}\,\sin\left(\frac{2|k|mx}{h}\right)$$

```
> probability:=P=(R0)^2;
```

$$probability := P = \frac{8|k|m\sin\left(\frac{2|k|mx}{h}\right)^2}{h\left(\frac{4A|k|m}{h} - \sin\left(\frac{4A|k|m}{h}\right)\right)}$$

```
> probability := {P =
8*abs(k)*m*sin(2*abs(k)*m/h*x)^2/(4*A*abs(k)*m-
sin(4*A*abs(k)*m/h)*h)};
```

$$probability := \left\{P = \frac{8|k|m\sin\left(\frac{2|k|mx}{h}\right)^2}{4A|k|m - \sin\left(\frac{4A|k|m}{h}\right)h}\right\}$$

```
>s1:=limit(8*abs(k)*m*sin(2*abs(k)*m/h*x)^2/(4*A*abs(k)*m
-sin(4*A*abs(k)*m/h)*h),k=0);
```

$$s1 := \frac{3\,x^2}{A^3}$$

```
> limit(s1,x=A);
```

$$\frac{3}{A}$$

4. This program to calculate time and n from continuity equation.

```
> 32*m*abs(k)*abs(k)*m*n*(n-1)*(t^(n-1))*(cos(abs(k)*m*A*n*2*(t^(n-1))/h)*sin(abs(k)*m*A*n*2*(t^(n-1))/h))=0;
```

$$32\,m^2\,|k|^2\,n\,(n-1)\,t^{(n-1)}\cos\left(\frac{2\,|k|\,m\,A\,n\,t^{(n-1)}}{h}\right)\sin\left(\frac{2\,|k|\,m\,A\,n\,t^{(n-1)}}{h}\right)=0$$

```
> R1 := solve({32*m^2*abs(k)^2*n*(n-1)*t^(n-1)*cos(2*abs(k)*m*A*n*t^(n-1)/h)*sin(2*abs(k)*m*A*n*t^(n-1)/h) = 0},{t});
```

$$R1 := \{t=0\},\ \left\{t=\mathbf{e}^{\left(\frac{\ln\left(1/4\,\frac{\pi\,h}{|k|\,m\,A\,n}\right)}{n-1}\right)}\right\},\ \left\{t=\mathbf{e}^{\left(\frac{\ln\left(-1/4\,\frac{\pi\,h}{|k|\,m\,A\,n}\right)}{n-1}\right)}\right\}$$

```
> R0 := solve({32*m^2*abs(k)^2*n*(n-1)*t^(n-1)*cos(2*abs(k)*m*A*n*t^(n-1)/h)*sin(2*abs(k)*m*A*n*t^(n-1)/h) = 0},{n});
```

$$R0 := \{n=0\},\ \{n=1\},\ \left\{n=\frac{\mathrm{LambertW}\left(\frac{1}{4}\,\frac{\ln(t)\,\pi\,t\,h}{|k|\,m\,A}\right)}{\ln(t)}\right\},$$

$$\left\{n=\frac{\mathrm{LambertW}\left(-\frac{1}{4}\,\frac{\ln(t)\,\pi\,t\,h}{|k|\,m\,A}\right)}{\ln(t)}\right\}$$

**************************************************************************

```
> ty:=proc(n);
> localresult;
> result: exp(ln(c/n)/(n-1));
> end;
```

$$ty := \mathbf{proc}(n)\ localresult\ ;\ result\ ;\ \exp(\ln(c/n)/(n-1))\ \mathbf{end\ proc}$$

```
> ty(n);
```

$$\mathbf{e}^{\left(\frac{\ln\left(\frac{c}{n}\right)}{n-1}\right)}$$

```
> R4 :=exp(ln(1/2*pi/(595275.3945*pi)/n)/(n-1));
```

$$R4 := \mathbf{e}^{\left(\frac{\ln\left(\frac{0.8399473666\,10^{-6}}{n}\right)}{n-1}\right)}$$

```
> r1:=ty(2/3)=5*10^17;
```

$$r1 := \frac{8}{27\,c^3} = 5000000000000000($$

```
> evalf(c);
```

$\{c = 0.8399473667\ 10^{-6}\}, \{c = -0.4199736833\ 10^{-6} + 0.7274157576\ 10^{-6}\ I\},$
$\{c = -0.4199736833\ 10^{-6} - 0.7274157576\ 10^{-6}\ I\}$

```
> r2:=ty(1/2);
```

$$r2 := \frac{1}{4\,c^2}$$

```
> tim:=1/(4*.8399473667e-6^2);
```

$$tim := 0.3543527952\ 10^{12}$$

When t=13.75 Gy ;

$\{c = -0.4402975634\ 10^{-6} - 0.7626177503\ 10^{-6}\ I\},$
$\{c = -0.4402975634\ 10^{-6} + 0.7626177503\ 10^{-6}\ I\}, \{c = 0.8805951268\ 10^{-6}\}$

5. This program to calculate (n)

```
> np:=proc(t);
> localresult;
> result:1/ln(t)*LambertW(.8399473666e-6*ln(t)*pi/pi*t);
> end;
```

*np* := **proc**(*t*)
*localresult*; *result*; LambertW(0.8399473666*10^(-6)×ln(*t*)×*t*)/ln(*t*)
**end proc**

```
> np(t);
```

$$\frac{\mathrm{LambertW}(0.8399473666\ 10^{-6}\ \ln(t)\ t)}{\ln(t)}$$

```
> np(10^12);
```

$$\frac{\mathrm{LambertW}(839947.3666\ \ln(1000000000000))}{\ln(1000000000000)}$$

```
> evalf(%);
```

$$0.5175266154$$

```
> np(5e17);
```

$$0.6666666668$$

```
> R1 := convert(.6720473448,'rational','exact');
> R0 := convert(.6720473448,'rational');
> np(10^(-36));
```

$$-\frac{\mathrm{LambertW}(-0.8399473666\ 10^{-42}\ \ln(1000000000000000000000000000000000000))}{\ln(1000000000000000000000000000000000000)}$$

```
> evalf(%);
```

$$0.8399473667\ 10^{-42}$$

```
> n=np(t);
```

$$n = \frac{\mathrm{LambertW}(0.8399473666\ 10^{-6}\ \ln(t)\ t)}{\ln(t)}$$

```
> np(.9999999999);
```

$$0.8399473665 \ 10^{-6}$$

```
> limit(np(t),t=1);
```

$$0.8399473666 \ 10^{-6}$$

```
> limit(np(t),t=infinity);
```

$$1.$$

```
> np(1);
Error, (in np) numeric exception: division by zero

> np(0);
Error, (in ln) numeric exception: division by zero

> np(5.39e-44);
```

$$0.4527316303 \ 10^{-49}$$

```
> np(.9);
```

$$0.7559526901 \ 10^{-6}$$

```
> np(1e-37);
```

$$0.8399473666 \ 10^{-43}$$

6. This program to calculate the scale factor (a)

```
> scale:=proc(n)
> localresult;
>
> result:A*exp(ln(1/2*pi/(595275.3945*pi)/n)/(n-1))^n;
> end;
```

*scale* := **proc**(*n*)

*localresult*; *result*; $A \times \exp(\ln(0.8399473666*10\text{^}(-6)/n)/(n-1))\text{^}n$

**end proc**

```
> scale(n);
```

$$A\left(e^{\left(\frac{\ln\left(\frac{0.8399473666\,10^{-6}}{n}\right)}{n-1}\right)}\right)^{n}$$

```
> scale(1/2);
```

$$595275.3973 \ A$$

```
> scale(2/3);
```

$$0.6299605277 \ 10^{12} \ A$$

```
> scale(.66667*10^(-6));
```

$$0.9999998460 \ A$$

```
> scale(.9);
```

$$0.1861718133 \ 10^{55} \ A$$

```
> scale(1.1);
```

$$0.5145804043 \ 10^{-67} \ A$$

*************************************************************

This to calculate $\dot{a}$ :

```
> vsc:=proc(n);
> localresult;
> result:A*n*exp(ln(1/2*pi/(595275.3945*pi)/n)/(n-1))^(n-
1);
> end;
```

$vsc := \textbf{proc}(n)$

$localresult;\ result;\ A \times n \times \exp(\ln(0.8399473666*10\hat{}(-6)/n)/(n-1))\hat{}(n-1)$

$\textbf{end proc}$

```
> vsc(n);
```

$$A\,n\left(e^{\left(\frac{\ln\left(\frac{0.8399473666\,10^{-6}}{n}\right)}{n-1}\right)}\right)^{(n-1)}$$

```
> vsc(1/2);
```

$$0.8399473625\ \ 10^{-6}\,A$$

```
> vsc(1/3);
```

$$0.8399473643\ \ 10^{-6}\,A$$

```
> vsc(2/3);
```

$$0.8399473647\ \ 10^{-6}\,A$$

```
> vsc(1/4);
```

$$0.8399473685\ \ 10^{-6}\,A$$

```
> vsc(1);
Error, (in vsc) numeric exception: division by zero

> vsc(0);
Error, (in vsc) numeric exception: division by zero

> vsc(.66667*10^(-7));
```

$$0.8399473664\ \ 10^{-6}\,A$$

```
> vsc(.99);
```

$$0.8399473667\ \ 10^{-6}\,A$$

```
> ad=n*exp(ln(1/2*pi/(595275.3945*pi)/n)/(n-1))^(n-1);
```

$$ad = n\left(e^{\left(\frac{\ln\left(\frac{0.8399473666\,10^{-6}}{n}\right)}{n-1}\right)}\right)^{(n-1)}$$

```
> vsc(.98);
```

$$0.8399473654\ \ 10^{-6}\,A$$

```
> limit(vsc(n),n=1);
```

$$0.8399473666\ \ 10^{-6}\,A$$

```
> limit(vsc(n),n=0);
```

$$0.8399473666\ \ 10^{-6}\,A$$

```
> limit(vsc(n),n=0);
```

$$0.8399473666\ \ 10^{-6}\,A$$

7. This program to calculate the acceleration

```
> acc:=proc(n);
> localresult;
>
> result:A*n*(n-1)*exp(ln(1/2*pi/(595275.3945*pi)/n)/(n-1))^(n-2);
> end;
```

$acc := \mathbf{proc}(n)$

$localresult;$

$result;$

$A \times n \times (n-1) \times \exp(\ln(0.8399473666*10^{(-6)}/n)/(n-1))^{(n-2)}$

$\mathbf{end\ proc}$

```
> acc(n);
```

$$A\,n\,(n-1)\left(e^{\left(\frac{\ln\left(\frac{0.8399473666\ 10^{-6}}{n}\right)}{n-1}\right)}\right)^{(n-2)}$$

```
> acc(1/2);
```

$$-0.1185185168\ 10^{-17}\,A$$

```
> acc(2/3);
```

$$-0.5599649062\ 10^{-24}\,A$$

```
> acc(.99);
```

$$-0.6108574821\ 10^{-615}\,A$$

```
> acc(1e-36);
```

$$-0.7055115763\ 10^{24}\,A$$

```
> acc(1);
Error, (in asc) numeric exception: division by zero

> acc(0);
Error, (in asc) numeric exception: division by zero

> acc(-1);
```

$$(0.9473319314\ 10^{-18} - 0.1539600709\ 10^{-8}\,I)\,A$$

```
> acc(1.1);
```

$$0.1246402218\ 10^{55}\,A$$

```
> acc(.8399473666e-6);
```

$$-0.8399466611\ 10^{-6}\,A$$

8. This program to calculate density:

```
> rho:=proc(n);
> localresult;
> result:3.61e9*(n/(exp(ln(1/2*pi/(595275.3945*pi)/n)/(n-1))))^2*(sin(1.87e6*n*exp(ln(1/2*pi/(595275.3945*pi)/n)/(n-1))^(n-1)))^2/A;
> end;
```

$\rho$ := **proc**($n$)

*localresult*;

*result*;

0.361*10^10×$n$^2×

sin(0.187*10^7×$n$×exp(ln(0.8399473666*10^(-6)/$n$)/($n$ − 1))^($n$ − 1))^2/

(exp(ln(0.8399473666*10^(-6)/$n$)/($n$ − 1))^2×$A$)

**end proc**

```
> rho(n);
```

$$\frac{0.361\ 10^{10}\, n^2 \sin\left(0.187\ 10^7\, n \left(\exp\left(\frac{\ln\left(\frac{0.8399473666\ 10^{-6}}{n}\right)}{n-1}\right)\right)^{(n-1)}\right)^2}{\left(\exp\left(\frac{\ln\left(\frac{0.8399473666\ 10^{-6}}{n}\right)}{n-1}\right)\right)^2 A}$$

```
> rho(2/3);
```

$$\frac{0.6417777634\ 10^{-26}}{A}$$

```
> rho(1/2);
```

$$\frac{0.7187460525\ 10^{-14}}{A}$$

```
> rho(.9);
```

$$\frac{0.7348242611\ 10^{-111}}{A}$$

```
> rho(.8399473668e-6);
```

$$\frac{0.00254689677}{A}$$

```
> rho(.4527316303e-49);
```

$$\frac{0.00254689678}{A}$$

…………………………………………………………………………………………..

Hubble parameter:

```
> localresult;
>
> result:n/(exp(ln(1/2*pi/(595275.3945*pi)/n)/(n-1)));
> end;
```

$hu$ := **proc**($n$) *localresult* ; *result* ; $n$/exp(ln(0.8399473666*10^(-6) /$n$)/($n$ − 1)) **end proc**

```
> hu(n);
```

$$\frac{n}{\exp\left(\frac{\ln\left(\frac{0.8399473666\ 10^{-6}}{n}\right)}{n-1}\right)}$$

```
> hu(1/2);
```

$$0.1411023144\ \ 10^{-11}$$

```
> hu(2/3);
```

$$0.1333333325\ 10^{-17}$$

```
> hh:=H=hu(n);
```

$$hh := H = \frac{n}{e^{\left(\frac{\ln\left(\frac{0.8399473666\ 10^{-6}}{n}\right)}{n-1}\right)}}$$

```
> limit(hh,n=0);
```

$$H = 0.8399473666\ 10^{-6}$$

```
> limit(hh,n=1);
```

$$H = \text{Float(undefined)}$$

```
> hu(.9);
```

$$0.4511678498\ 10^{-60}$$

```
> hu(.001);
```

$$0.6618041261\ 10^{-6}$$

.......................................................................................................

redshift:

```
> z:=proc(n0,n);
> localresult;
>
> result:(exp(ln((.8399473667e-6)/n0)/(n0-
1)))^n0/(exp(ln((.8399473667e-6)/n)/(n-1)))^n;
> end;
```

$z :=$ **proc**$(n0, n)$

*localresult*;

*result*;

exp( ln( 0.8399473667*10^(-6)/n0)/(n0 − 1))^n0/ **end proc**

exp( ln( 0.8399473667*10^(-6)/n)/(n − 1))^n

```
> z(n0,n);
```

$$\frac{\left(e^{\left(\frac{\ln\left(\frac{0.8399473667\ 10^{-6}}{n0}\right)}{n0-1}\right)}\right)^{n0}}{\left(e^{\left(\frac{\ln\left(\frac{0.8399473667\ 10^{-6}}{n}\right)}{n-1}\right)}\right)^{n}}$$

```
> z(2/3,1/2);
```

$$0.1058267368\ 10^{7}$$

```
> z(2/3,1/3);
```

$$0.1000000003\ 10^{10}$$

```
> z(2/3,.8399473666e-6);
```

$$0.6299605277\ 10^{12}$$

```
> z(.6962745918,.8399473666e-6);
```

$$0.3697571388\ 10^{14}$$

9. Imaginary scale factor:

```
> scale:=proc(n);
> localresult;
>
> result:(-3.139999999-5.438639534*I)*(exp(ln((-
.4402975634e-6+.7626177503e-6*I)/n)/(n-1)))^n;
> end;
```

*scale* := **proc**(*n*)

*localresult*;

*result*;

( − 3.139999999− 5.438639534*I) * (exp(ln(( − 4.402975634

·10^-7 + 7.626177503 10^-7*I)/*n*)/(*n* − 1)))^*n*

**end proc**

```
> scale(n);
```

$$(-3.139999999 - 5.438639534\,I)\left(e^{\frac{\ln\left(\frac{-4.402975634\ 10^{-7} + 7.626177503\ 10^{-7}\ I}{n}\right)}{n-1}}\right)^{n}$$

```
> scale(1/2);
```

$$1.782885185\,10^{6} - 3.088047721\,10^{6}\,I$$

```
> scale(2/3);
```

$$-1.799677035\,10^{12} - 3.117132067\,10^{12}\,I$$

```
> (scale(2/3)/scale(1/2))^3*4/3*3.14159/5.67e-8;
```

$$-7.598341041\,10^{25} - 5.226702138\,10^{16}\,I$$

.............................................................................................

10. This program to calculate imaginary $\dot{a}$ :

```
> vsc:=proc(n); > vsc:=proc(n);
> localresult;
>
> result:(-3.139999999-5.438639534*I)*n*exp(ln((-
.4405626353e-6+.7630768683e-6*I)/n)/(n-1))^(n-1);
> end;
```

*vsc* := **proc**(*n*)

*localresult*;

*result*;

( − 3.139999999− 5.438639534*I) * *n* * (exp(ln(( − 4.40562635

·10^-7 + 7.630768683 10^-7*I)/*n*)/(*n* − 1)))^(*n* − 1)

**end proc**

```
> vsc(n);
```

$$(-3.139999999 - 5.438639534 I)\, n \left( e^{\frac{\ln\left(\frac{-4.405626353\ 10^{-7} + 7.630768683\ 10^{-7}\, I}{n}\right)}{n-1}} \right)^{n-1}$$

```
> vsc(1/2);
```

$$-0.000005533466694 + 3.018166856\ 10^{-15}\, I$$

```
>
> vsc(2/3);
>
```

$$-0.000002766733355 - 0.000004792122748 I$$

```
> vsc(.8399473666e-6);
```

$$0.000005533466698 - 2.956574995\ 10^{-15}\, I$$

```
> vsc(.4527316303e-49);
```

$$0.000005533466678 - 2.322534299\ 10^{-15}\, I$$

```
> vsc(.9);
```

$$-0.000001709935257 - 0.000005262639581 I$$

```
> limit(vsc(n),n=0);
```

$$0.000005533466698 - 1.531191815\ 10^{-16}\, I$$

```
> limit(vsc(n),n=1);
```

$$-0.000002766733348 - 0.000004792122730 I$$

..........................................................................................

11. This program to calculate imaginary *ä*

```
> ac:=proc(n);
> localresult;
>
> result:(-3.14159-5.441393498*I)*n*(n-1)*exp(ln((-
.4405626353e-6+.7630768683e-6*I)/n)/(n-1))^(n-2);
> end;
```

*ac* := **proc**(*n*)
*localresult*;
*result*;
( − 3.14159 − 5.441393498*I) * *n* * (*n* − 1) * (exp(ln((
− 4.405626353 10^-7 + 7.630768683 10^-7 * I) / *n*) / (*n* − 1)))
^(*n* − 2)
**end proc**

```
> ac(n);
```

$$(-3.14159 - 5.441393498 I)\, n\, (n - 1) \left( e^{\frac{\ln\left(\frac{-4.405626353\ 10^{-7} + 7.630768683\ 10^{-7}\, I}{n}\right)}{n-1}} \right)^{n-2}$$

```
> ac(1/2);
```

$$-4.298257920\ 10^{-18} - 7.444801082\ 10^{-18}\, I$$

```
> ac(2/3);
```

$2.130358351 10^{-24} + 3.689888924 10^{-24} I$

```
> ac(.8399473666e-6);
```

0.000002903846946 – 0.000005029590028 I

```
>
> ac(.4527316303e-49);
```

$5.387459000 10^{37} - 9.331352738 10^{37} I$

```
> limit(ac(n),n=0);
```

*Float*(*undefined* )

```
> limit(ac(n),n=1);
```

*Float*(*undefined* )

12. This program to calculate imaginary H

```
> hu:=proc(n);
> localresult;
> result:n/exp(ln((-.4405626353e-6+.7630768683e-
6*I)/n)/(n-1));
> end;
```

*hu* := **proc**(*n*)
*localresult*;
*result*;
*n*/exp( ln( -0.4405626353*10^(-6) + 0.7630768683*10^(-6)*I/*n* )/( *n* – 1 ) )
**end proc**

```
> hu(n);
```

$$\frac{n}{e^{\left(\frac{\ln\left(\frac{-0.4405626353\ 10^{-6} + 0.7630768683\ 10^{-6}\ I}{n}\right)}{n-1}\right)}}$$

```
> hu(1/2);
```

$-0.7763817423\ 10^{-12} - 0.1344732621\ 10^{-11}\ I$

```
> hu(2/3);
```

$0.1539201547\ 10^{-17} + 0.2801983311\ 10^{-26}\ I$

```
> hu(.9);
```

$-0.3640524881\ 10^{-60} + 0.6305574119\ 10^{-60}\ I$

```
> hu(.8399473666e-6);
```

$-0.4405639950\ 10^{-6} + 0.7630761242\ 10^{-6}\ I$

```
> hu(.71619913179);
```

$0.4899236296\ 10^{-21} + 0.9545864800\ 10^{-21}\ I$

```
> hu(1/3);
```

$-0.1432574160\ 10^{-8} + 0.8449225532\ 10^{-18}\ I$

13. This program to calculate imaginary density

```
> rho:=proc(n);
> localresult;
>
> result:(3/8/(-.1276082296+.2164992928*I)/3.14/6.672e-11)*(n/exp(ln((-.4402975634e-6+.7626177503e-6*I)/n)/(n-1)))^2*(sin((3.14/2/(-.4402975634e-6+.7626177503e-6*I)*n*(exp(ln((-.4402975634e-6+.7626177503e-6*I)/n)/(n-1)))^(n-1)))^2);
> end;
```

ρ := **proc**($n$)
  *localresult*;
  *result*;
  (-0.3616687599*10^10-0.6136048669*10^10*I)×n^2×sin(
    (-891442.5893-0.1544023857*10^7*I)×n×
    exp(ln(-0.4402975634*10^(-6) + 0.7626177503*10^(-6) ×I/n)/(n − 1))^
    (n − 1))^2/
    exp(ln(-0.4402975634*10^(-6) + 0.7626177503*10^(-6) ×I/n)/(n − 1))^2

**end proc**

```
> rho(n);
```

$$\frac{(-0.3616687599\,10^{10} - 0.6136048669\,10^{10}\,I)\,n^2\,\sin\left((891442.5893 + 0.1544023857\,10^{7}\,I)\,n\left(e^{\left(\frac{\ln\left(\frac{-0.4402975634\,10^{-6} + 0.7626177503\,10^{-6}\,I}{n}\right)}{n-1}\right)}\right)^{(n-1)}\right)^2}{\left(e^{\left(\frac{\ln\left(\frac{-0.4402975634\,10^{-6} + 0.7626177503\,10^{-6}\,I}{n}\right)}{n-1}\right)}\right)^2}$$

```
> rho(1/2);
```

$$0.1713113633\ 10^{-13} - 0.1542238948\ 10^{-15}\,I$$

```
> rho(2/3);
```

$$0.5045186728\ 10^{-25} - 0.3943643036\ 10^{-25}\,I$$

14. Imaginary redshift:

```
> z:=proc(n0,n);
> localresult;
>
> result:(n/n0)*(exp(ln((-.4402975634e-6+.7626177503e-6*I)/n0)/(n0-1)))^n0/(exp(ln((-.4402975634e-6+.7626177503e-6*I)/n)/(n-1)))^n;
```

```
> end;
```

$z :=$ **proc**$(n0, n)$

*local result;*

*result;*

$n \times \exp(\ln(-0.4402975634*10\text{^}(-6) + 0.7626177503*10\text{^}(-6)*I/n0)/(n0-1))\text{^}n0 / (n0 \times \exp(\ln(-0.4402975634*10\text{^}(-6) + 0.7626177503*10\text{^}(-6)*I/n)/(n-1))\text{^}n)$

**end proc**

```
> z(n0,n);
```

$$\frac{n\left(e^{\left(\frac{\ln\left(\frac{-0.4402975634\,10^{-6}+0.7626177503\,10^{-6}\,I}{n0}\right)}{n0-1}\right)}\right)^{n0}}{n0\left(e^{\left(\frac{\ln\left(\frac{-0.4402975634\,10^{-6}+0.7626177503\,10^{-6}\,I}{n}\right)}{n-1}\right)}\right)^{n}}$$

```
> z(2/3,1/2);
```

$$378531.8851 - 655636.4568\, I$$

```
> z(2/3,1/3);
```

$$0.2326813813\; 10^{9} - 0.4030159730\; 10^{9}\, I$$

```
> z(2/3,.8399473666e-6);
```

$$721249.4858 - 1.267934611\, I$$

```
> z(.6962745918,.8399473666e-6);
```

$$0.3639034496\; 10^{8} + 0.1653402114\; 10^{8}\, I$$

```
> z(1/2,.8399473666e-6);
```

$$280831.8725 + 486417.0481\, I$$

```
> limit(z(n0,n),n0=1);
```

$$0.$$

```
> limit(z(n0,n),n0=0);
```

$$\frac{1}{\left(e^{\left(\frac{\ln\left(\frac{-0.4451048383\,10^{-6}+0.7709441947\,10^{-6}\,I}{n}\right)}{n-1.}\right)}\right)^{n}}$$